\documentclass[aps, reprint, groupedaddress, pre, showpacs]{revtex4-1}

\newcommand{\textupperscript}[1]{$^{\text{#1}}$}
\newcommand{\textunderscript}[1]{$_{\text{#1}}$}
\newcommand{\ca}{Ca\textupperscript{2+} }
\newcommand{\ipr}{IP\textunderscript{3}R }
\newcommand{\ip}{IP\textunderscript{3} }

\usepackage{graphicx}

\usepackage{commath}

\usepackage{color}

\usepackage{caption}

\usepackage[hidelinks,colorlinks=true,
linkcolor=black,
urlcolor=blue,
citecolor=black]{hyperref}

\begin{document}


\title{A phenomenological cluster-based model of Ca$^\text{2+}$ waves and oscillations for Inositol 1,4,5-trisphosphate receptor (IP\textunderscript{3}R) channels}


\author{Svitlana Braichenko}
\altaffiliation{Corresponding author}
\email{braichenko.svitlana@gmail.com}
\author{Atul Bhaskar}
\affiliation{Faculty of Engineering and the Environment, University of Southampton, SO17 1BJ, Southampton, UK}

\author{Srinandan Dasmahapatra}
\affiliation{School of Electronics and Computer Science, University of Southampton, SO17 1BJ, Southampton, UK}


\date{\today}

\begin{abstract}
\end{abstract}


\begin{abstract}
Clusters of IP\textunderscript{3} receptor channels in the membranes of the endoplasmic reticulum (ER) of many non-excitable cells release calcium ions in a cooperative manner giving rise to dynamical patterns such as Ca\textupperscript{2+} puffs, waves, and oscillations that occur on multiple spatial and temporal scales. We introduce a minimal yet descriptive reaction-diffusion model of IP\textunderscript{3} receptors for a saturating concentration of IP\textunderscript{3} using a principled reduction of a detailed Markov chain description of individual channels. A dynamical systems analysis reveals the possibility of excitable, bistable and oscillatory dynamics of this model that correspond to three types of observed patterns of calcium release -- puffs, waves, and oscillations respectively. We explain the emergence of these patterns via a bifurcation analysis of a coupled two-cluster model, compute the phase diagram and quantify the speed of the waves and period of oscillations in terms of system parameters. We connect the termination of large-scale \ca release events to \ip unbinding or stochasticity.  
\end{abstract}

\maketitle


\section{\label{sec:INTRO}INTRODUCTION}
Ionic calcium (Ca\textupperscript{2+}) is a second messenger involved in many processes such as fertilization, cell proliferation, differentiation, embryonic development, secretion, muscular contraction, immune response, brain functions, chemical sensing, light transduction, etc.~\cite{Berridge1997a,Mikoshiba2007}. While many types of \ca release channels are implicated in these processes, in this paper we are concerned with the dynamics of Ca\textupperscript{2+} released into the cytosol from inositol 1,4,5-triphosphate receptor~(IP\textunderscript{3}R) channels~\cite{Mikoshiba2007} that are triggered by \ip secondary messenger molecules. IP\textunderscript{3}R channels are usually present in non-excitable cells and are mainly localized in the endoplasmic reticulum~(ER) membrane where they are believed to form tight clusters to enable Ca\textupperscript{2+} signaling. Experimental observations~\cite{Taufiq-Ur-Rahman2009,Smith2014} suggest that the distribution of clusters varies by cell type. For example, SH-SY5Y neuroblastoma cells contain around $4$ to $6$ channels per cluster (a few contain up to $10$ channels), HeLa cells clusters on average contain $2$ to $3$ channels and \textit{Xenopus} oocytes clusters contain around $20$ channels each~\cite{Dickinson2012}. However, theoretical and experimental investigations of clustering and the mechanisms by which it regulates Ca\textupperscript{2+} signaling are incomplete and is the subject of ongoing research. 

Ca\textupperscript{2+} signaling events corresponding to release from Ca\textupperscript{2+} channels~\cite{Berridge1997a} constitute a multiscale hierarchy consisting of three distinct types of events taking place at different spatial and temporal scales involving individual channels, clusters of channels and groups of channel clusters. The events occurring at the smallest scale ($\sim10\,\text{nm}$ and $\sim10\,\text{ms}$) are called \textit{blips}. They are the building blocks of signals on larger scales. Blips occur when Ca\textupperscript{2+} is released from a single channel into the cytosol. Coordinated \ca release from a cluster of channels -- a collection of co-occurring blips --  is called a puff and appears at the second level of the hierarchy ($\sim100\,\text{nm}$ and $\sim100\,\text{ms}$).  This cooperation between signaling events is regulated by Ca\textupperscript{2+} in a concentration-dependent feedback mechanism. Low Ca\textupperscript{2+} concentrations (maximum activity for $200-500$ nM) diffuse around the cytosol and cause further release from neighboring channels, whereas Ca\textupperscript{2+} in high cytosolic concentrations inhibits further Ca\textupperscript{2+} release. This feedback mechanism is called Ca\textupperscript{2+} induced Ca\textupperscript{2+} release (CICR). The final level in the hierarchy ($\sim1\,\mu \text{m}$ and $\sim1\,\text{s}$) is associated with Ca\textupperscript{2+} \textit{spikes}, \textit{waves}, and \textit{oscillations}. Spikes correspond to local or global transient releases from a group of clusters. Ca\textupperscript{2+} waves are formed as sequential releases travel from excited clusters to neighboring ones. Ca\textupperscript{2+} oscillations are repetitive Ca\textupperscript{2+} spikes or waves.  All the events on the highest hierarchical level that we consider (spikes, waves, oscillations) consist of the coordinated release of Ca\textupperscript{2+} puffs from a group of clusters.  

Ca\textupperscript{2+} oscillations are experimentally observed in various types of cells~\cite{Berridge2007}. Many models associate the occurrence of Ca\textupperscript{2+} oscillations with the IP\textunderscript{3} regulation~\cite{DeYoung1992,Li1994,Politi2006,Smedler2014,Berridge2007}. Some studies connect the emergence of Ca\textupperscript{2+} oscillations with the oscillating level of IP\textunderscript{3} triggered by binding to various enzymes, such as PLC (phospholipase C) or IP\textunderscript{3} kinase~\cite{Politi2006,Berridge2007}. Others report that Ca\textupperscript{2+} oscillations may occur even if the IP\textunderscript{3} concentration is non-oscillatory~\cite{DeYoung1992,Li1994}. Various studies explain oscillatory Ca\textupperscript{2+} behavior as a sequence of stochastic spikes~\cite{Falcke2003,Skupin2008,Thurley2011}. Apart from oscillations and spikes  propagating wavefronts have also been much studied.  These are implicated, for instance, in oocyte fertilization, and are generated when a sperm cell makes contact with an oocyte.

There is an extensive literature on the deterministic modeling of \ca front propagation. References~\cite{Kupferman1997, DeYoung1992, Atri1993, Li1994, Sneyd1998, Dupont1994} consider the averaged \ca dynamics from a well-mixed ER membrane without addressing the clustering of channels. Channel clusters, including inhomogeneous cluster distributions~\cite{Calabrese2010}, are considered in deterministic~\cite{Keizer1998, Sneyd1997, Dawson1999} and stochastic models~\cite{Keizer1998, Coombes2003} of \ca wave propagation. The current paper studies the deterministic properties of \ca waves and oscillations occurring in the ER membrane containing multiple clustered channels. The models in~\cite{Sneyd1997, Dawson1999} inject a burst of \ca ions at each cluster which diffuses between clusters.  While the burst size is represented by a parameter in their models, we choose to represent the release mechanism by a reduced version of the DYK~\cite{DeYoung1992} model proposed by R{\"{u}}diger~\cite{Rudiger2014}.  This gives our model access to single channel characteristics~\cite{DeYoung1992} and cluster properties~\cite{Rudiger2014} providing more mechanistic insight into the link between single-channel properties and features of waves and oscillations at higher levels in the spatial hierarchy.    We use  dynamical systems analysis to explore the different dynamical regimes at the single-cluster level and relate these properties to larger spatial scales. Thus, unlike the fire-diffuse-fire model~\cite{Sneyd1997, Dawson1999} we can study the characteristics of global \ca release events such as waves and oscillations,  including their inhibition and termination, using channel and cluster parameters.  We also build on the results of a detailed hybrid stochastic-deterministic single-cluster study~\cite{Ruckl2015} and its extension on a grid of clusters~\cite{Ruckl2016} that show different durations of \ca events depending on the \ip level.

In detail, we introduce a spatial model of Ca\textupperscript{2+} waves and oscillations building upon qualitative insights from dynamical systems theory applied to reduced, few variable models~\cite{Li1994,Rudiger2014} of \ipr  channel subunits~\cite{DeYoung1992} integrated into channels and heterogeneous clusters of channels~\cite{Ruckl2015}. We assume the IP\textunderscript{3} concentration to be sustained at a high level so that it increases the probability of the occurrence of Ca\textupperscript{2+} waves and oscillations, extending the observation of excitability of channel clusters in~\cite{Rudiger2014}. Further, we account for variability of [IP\textunderscript{3}] levels by including the \ip unbinding. The extended model also gives insight on the numbers of activatable channels (all four subunits has \ip bound~\cite{Taylor2016}) in agreement with~\cite{Ruckl2015,Ruckl2016}. The model we propose is of a deterministic reaction-diffusion type where diffusion of \ca between clusters of channels smooths the  \ca releases from individual activated clusters modeled by reaction dynamics. Here we study the dependence of the velocity and period of waves and oscillations on the diffusion coefficient, equilibration rate and a number of channels in each cluster. We observe a curious spatially alternating wave pattern in a regime of parameter space that is reminiscent of the occurrence of intracellular spatially organized \ca alternans \cite{Qu2016}. The termination of large-scale events such as waves and oscillations has been a topic of considerable discussion~\cite{Groff2008,Hinch2004,Wiltgen2014,Laver2013,Rudiger2012}. We provide evidence for two different mechanisms -- a stochastic termination and a deterministic \ip unbinding -- that can be accommodated in our models in order to account for the decay of \ca excitations in simulation traces.

\section{\label{sec:METH}METHODS}

A kinetic model of channel activation and deactivation is the basis for models of \ca dynamics, and many models~\cite{Li1994,Rudiger2014} have been proposed that reduce the detailed kinetic scheme laid out by De Young and Keizer -- the so-called DYK model~\cite{DeYoung1992} as a starting point. Our approach to simplifying the complexity of the problem is based on several basic models briefly discussed in this section. The assumption of a well-mixed membrane where spatial effects are ignored is a shared limitation of these models \cite{DeYoung1992,Li1994}. The spatial extension of the three-state model~\cite{Rudiger2014} is proposed in the last subsection. For further details please refer to the Appendices and Supporting Material. 

\subsection{\label{subsec:DYK}DYK and derived models of \ca signaling}

\begin{figure}[h!]
	\includegraphics{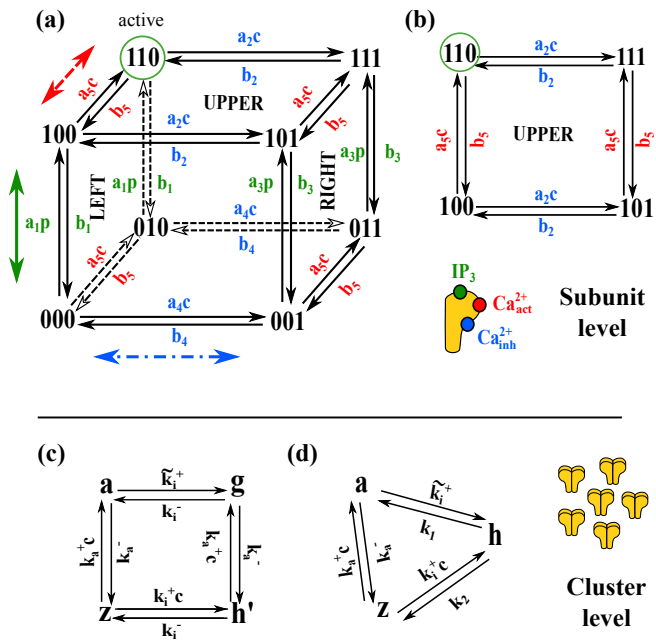}
	\caption{\label{fig:DYKRud} (Color online) \textbf{(a) DYK cube.} Schematic diagram of the DYK eight-state gating model of the activation of a  single subunit of an IP$_\text{3}$R channel; the eight corners of the schematic will be referred as the DYK cube~\cite{DeYoung1992}. Each vertex of the cube directly corresponds to a specific state of a subunit labeled with numbers $ijk$, where $i$ corresponds to the IP\textunderscript{3} activating site, $j$ to the Ca\textupperscript{2+} activating site and $k$ to the Ca\textupperscript{2+} inhibiting site. Each binding site can be either unbound ($i, j, k = 0$), or bound ($i, j, k = 1$). The \textit{green arrow} indicates binding (unbinding) of \ip to the binding site while \textit{red dashed} and \textit{blue dot-dashed arrows} correspond to \ca activating and inhibiting transitions respectively. The transitions between the states are governed by the second-order ($a_lp$ or $a_lc$, where $p=$ [IP\textunderscript{3}], $c=$ [Ca\textupperscript{2+}]), and the first-order ($b_l$) rate constants, where $l=\overline{1,5}$. \textbf{(b) Upper DYK plane.} When $p$ is high, the most of the transitions take place in the upper plane of the cube. \textbf{(c) Four-state model.} The scheme of the model~\cite{Rudiger2014} for high $p$ at the level of a cluster of channels. The fractions of the states in a cluster are marked as $a,g,z,h'$ instead of $110, 111, 100, 101$ for subunits. The transition rates between channel states are $k^\pm_{a,i}$ and $\tilde{k}^+_i=k^+_ic_s$ to indicate the elimination of detailed balance in a cluster of channels~\cite{Rudiger}. \textbf{(d) Three-state model.} The final model includes a compound state $h$ formed from the states $g$ and $h'$ with the effective rates $k_{1,2}$.}
\end{figure}

The DYK model~\cite{DeYoung1992} is one of the first models which successfully accounts for key empirical findings of IP\textunderscript{3}R channels. In the model each channel consists of four subunits, at least three of which need to be activated for channel opening, reflecting the three conductance levels experimentally observed in~\cite{Bezprozvanny1991}.  Each subunit has 3 binary state variables  $i,j,k\in\{0,1\}$, corresponding to 3 binding sites -- for IP\textunderscript{3}, activating \ca and inhibiting \ca -- with a state transition diagram shown in Fig.~\ref{fig:DYKRud}~(a) and described in its caption. The fraction of subunits in a given state ($ijk$) is represented by the variable $x_{ijk}$, such that they sum up to unity, i.e. $\sum_{i=0}^{1}\sum_{j=0}^{1}\sum_{k=0}^{1}x_{i,j,k}=1$. The presence of IP\textunderscript{3} at each channel enhances Ca\textupperscript{2+} release, while calcium either amplifies or inhibits further Ca\textupperscript{2+} release (called calcium-induced calcium release, CICR).  

The DYK model assumes that all 4-subunit channels are ``well-mixed" in the membrane and dispenses with spatial effects such as clustering. The complexity of the state space of the model makes it computationally difficult to obtain properties of the spatial distribution of channel clusters in the ER membrane. Hence, multiple attempts at reducing its complexity have been made in the last few decades, the majority of which share the common feature of IP\textunderscript{3}-dependent activation and CICR. 

In one such reduced model \cite{Li1994}, the variables corresponding to the DYK cube [Fig.~\ref{fig:DYKRud}~(a)] are dispensed with by assuming that activation by IP\textunderscript{3} and Ca\textupperscript{2+} are fast, compared with the slow inhibition by Ca\textupperscript{2+}.  The state indices $(i,j,0)$ and $(i,j,1)$ distinguish the two states of the site where the inhibiting \ca binds.  The fractions $x_{ij0}$ and $x_{ij1}$ of states with variable \ip and activating \ca binding sites on the left and the right faces of the DYK cube [Fig.~\ref{fig:DYKRud}~(a)] are slow variables in the slow-fast reduction.  The pairs of fractions $(x_{0j0}, x_{1j0})$ and $(x_{0j1}, x_{1j1})$ are determined from the steady state of the fast \ip binding and unbinding processes. The dynamics of the reduced model is governed by the compound dynamics of the left-right faces and Ca\textupperscript{2+}.  

\subsection{Single-cluster models}
Another reduction of the dynamics of a group of DYK subunits is introduced in \cite{Rudiger2014} where, in contrast to \cite{Li1994}, dynamical variables capture the collective behavior of a cluster of \ipr channels. This model shows that for large values of \ip the dynamics is excitable, and explains the occurrence of \ca puffs.   As this model is the starting point of our analysis we present the reduction of the \ip dynamics explicitly in Appendix~\ref{A1} for completeness. Here we discuss its fundamental assumptions. The reduction scheme follows  a similar slow-fast approach \cite{Li1994} but for the case of high $p=$ [IP\textunderscript{3}]. This assumption leads to fast transitions with rates $a_{1,3}p$ [Fig.~\ref{fig:DYKRud}~(a)] and all channel subunits are forced to have the \ip binding site occupied leaving only $1jk$ states to be dynamical (the upper plane of the DYK cube in Fig.~\ref{fig:DYKRud}~(a)).  The dynamical variables of the four subunits shown in Fig.~\ref{fig:DYKRud}~(b) are gathered together to introduce cluster variables $a$, $g$, $h'$, and $z$.  These represent the fractions of channels within a cluster in the corresponding states shown in Fig.~\ref{fig:DYKRud}~(c). The transition rates $k^-_a$, $k^\pm_i$ are derived from the parameters in the DYK model~\cite{DeYoung1992}. The transition rate $k_a^+$ is obtained from the requirement that at least three subunits need to be activated to open a channel (\ref{eq:ka+}). The numerical values of the coefficients \cite{Rudiger2014} are provided in Table~\ref{tab:RudOrig}.  A further simplification -- corresponding to the passage from Fig.~\ref{fig:DYKRud}~(c) to (d) --  is introduced \cite{Rudiger2014} by replacing the pair of states $(g,h')$ representing \ip-bound, inactivating \ca-bound states with a compound state $h$ whose kinetics has effective rates $k_1$ and $k_2$ [(\ref{eq:k1}), (\ref{eq:k2})].  This reduces the cluster dynamics to a two-state model as the subunit fractions within a cluster sum to unity, $z=1-a-g-h'=1-a-h$. A significant modeling assumption is made whereby the rates $\widetilde{k}_i^+$ and $k_i^+c$ in Fig.~\ref{fig:DYKRud}~(c) are taken to be different~\cite{Rudiger}, in contrast to the equal rates $a_2c$ on the upper plane of Fig.~\ref{fig:DYKRud}~(a) made in~\cite{DeYoung1992}, a condition that implies detailed balance.

The three-state model of a cluster of \ip channels thus obtained exhibits excitable and bistable dynamical regimes~\cite{Rudiger2014}.  In our study, we modify the model slightly, by setting the parameters for a single cluster of channels to fit the open probability distribution of \ip channels in the presence of \ca in \textit{Xenopus} oocytes~\cite{Ruckl2015}. This leads to changes in the transition rates $a_2$ and $b_2$ as shown in Tab.~\ref{tab:RudOrig}.  Consequently, an oscillatory regime emerges in addition to the excitable and bistable regimes found in~\cite{Rudiger2014}.  Later in this paper, we spatially extend this model in a cluster-based manner to study the interplay between the different regimes of clusters in producing wave and oscillation dynamics in a membrane.
	 	
\begin{table}[b]
	\caption{\label{tab:RudOrig}Parameters used in the model in~\cite{Rudiger2014} and the single cluster model of this paper.}
	\begin{ruledtabular}
		\begin{tabular}{c c c}
			Parameter & R{\"{u}}diger~\cite{Rudiger2014} & Present study \\	
			\hline
			Ca\textupperscript{2+} activation binding &  &  \\
			rate $a_5,\,\mu \text{M}^{-1}\times \text{s}^{-1}$ & $100$ & $100$ \\
			Ca\textupperscript{2+} activation unbinding &  &  \\
			rate ($b_5, k_a^-)\footnotemark[1],\,\text{s}^{-1}$ & $20$ & $20$ \\
			Ca\textupperscript{2+} inhibition binding &  & \\
			rate ($a_2, k_i^+)\footnotemark[1],\,\mu \text{M}^{-1}\times \text{s}^{-1}$ & $0.1$ & $0.02$\footnotemark[1] \\
			Ca\textupperscript{2+} inhibition unbinding &  &  \\
			rate ($b_2, k_i^-)\footnotemark[1],\,\text{s}^{-1}$ & $1.7$ & $1.56$\footnotemark[2] \\
			Local [Ca\textupperscript{2+}] at opened &  &  \\
			channel $\tilde{k}^+_i=k^+_ic_s,\,\text{s}^{-1}$ & $30$ & $6$\footnotemark[2] \\
			Rest level of [Ca\textupperscript{2+}] &  &  \\
			$c_0,\,\mu \text{M}$ & $0.025$ & $0.025-0.6$ \\
			Channel coupling constant &  &  \\ 
			$\alpha ,\,\mu \text{M}$ & $0.74$ & $0.74$ \\
			Number of activatable &  &  \\
			channels $N$ & $9$ or $25$ & $5$ or $6$ \\
			Equilibration rate $\lambda,\,\text{s}^{-1}$ & $10^3$ & $10^3$ \\
			Characteristic of the step $\epsilon$ & $0.1$ & $0.1$ \\
		\end{tabular}
	\end{ruledtabular}
    \footnotetext[1]{ Here we show the transition rates for subunit and channel as shown in Figs.~\ref{fig:DYKRud}~(b) and~(c) respectively.  These rates are assumed to be the same in this study. }
	\footnotetext[2]{These values are determined from patch-clamp data on the probability distribution for channel opening typical for \textit{Xenopus} oocytes at high [IP\textunderscript{3}]~\cite{Ruckl2015}}
\end{table}

The dynamics of a cluster is governed by the set of dynamical variables $x(t) =$ $\{a(t)$,$h(t)$,$c(t)\}$ corresponding to the fraction in the cluster of opened channels $a(t)$, inhibited channels $h(t)$, and the cytosolic Ca\textupperscript{2+} concentration $c(t)$. The dynamical variables in the phase space, as shown in Fig.~\ref{fig:DYKRud}~(d), are governed by a system of coupled ordinary differential equations~\cite{Rudiger2014} $\dot{x}=F(x(t))$  with appropriate initial conditions:
\begin{eqnarray}
\frac{\dif a}{\dif t} & = & k_{a}^{+}(c)c(1-a-h)-k_{a}^{-}a+k_{1}(c)h-\tilde{k}_{i}^{+}a \nonumber \\
&&\triangleq f(a,h,c),\label{eq:active}\\
\frac{\dif h}{\dif t} & = & k_{i}^{+}c(1-a-h)-k_{1}(c)h-k_{2}(c)h+\tilde{k}_{i}^{+}a \nonumber \\
&&\triangleq g(a,h,c).\label{eq:closed}
\end{eqnarray}
The rates $k_a(c)$, $k_{1}(c)$ and $k_{2}(c)$ are given in eqs.~(\ref{eq:ka+}), (\ref{eq:k1}) and (\ref{eq:k2}), respectively~\cite{Rudiger2014, Rudiger2012}. It is assumed in~\cite{Rudiger} that after a blip \ca levels at a channel pore remain high while neighboring channels in a cluster relax to levels around $c_d$. According to \cite{Shuai2008} [Ca\textupperscript{2+}] at a nanodomain drops rapidly from several hundreds of $\mu$M to a few $\mu$M. Thus, at the timescales of the dynamics studied in~\cite{Rudiger2014} and in our paper, the calcium-dependent inactivation rate is taken to be frozen at $\tilde{k}_i^+=k_i^+c_s$ as shown in the transition scheme Fig.~\ref{fig:DYKRud}~(c), (d).  The dynamics of the $c$ variable accounts for homeostatic mechanisms such as pumps and buffers in the simplest way, whereby \ca levels relax back to a steady-state level $c_d(a)$ at rate $\lambda$:
\begin{equation}
\frac{\dif c}{\dif t}=-\lambda[c-c_{d}(a)],\label{eq:CaRel}
\end{equation} 
where $\lambda=10^3\,\text{s}^{-1}$ \cite{Rudiger2014} sets the fast time-scale so that $c_d(a)$ is slaved to the dynamics of $a(t)$. The activation barrier for a cluster of channels is modeled separately by setting a threshold below which the cytosolic level of calcium is $c_0$ and above which the concentration of \ca is approximately linear in the number of opened channels $Na$:
\begin{equation}\label{eq:cd}
c_{d}(a)=c_{0}+0.5\alpha Na\left\{1+\tanh[(Na-1)/\epsilon]\right\},
\end{equation}   
where $\epsilon$ sets the scale for smoothing the transition from zeroth to first order dependence on opening and $\alpha$ is a constant defining the strength of the coupling between channels in a cluster. Such a step is proposed in \cite{Rudiger2014} to avoid \ca release from the inactive clusters where the number of opened channels $Na<1$. This modeling step is necessary because $a$ and $h$ denoting fractions of opened or inhibited channels are continuous variables.  

\begin{figure}[h]
	\includegraphics{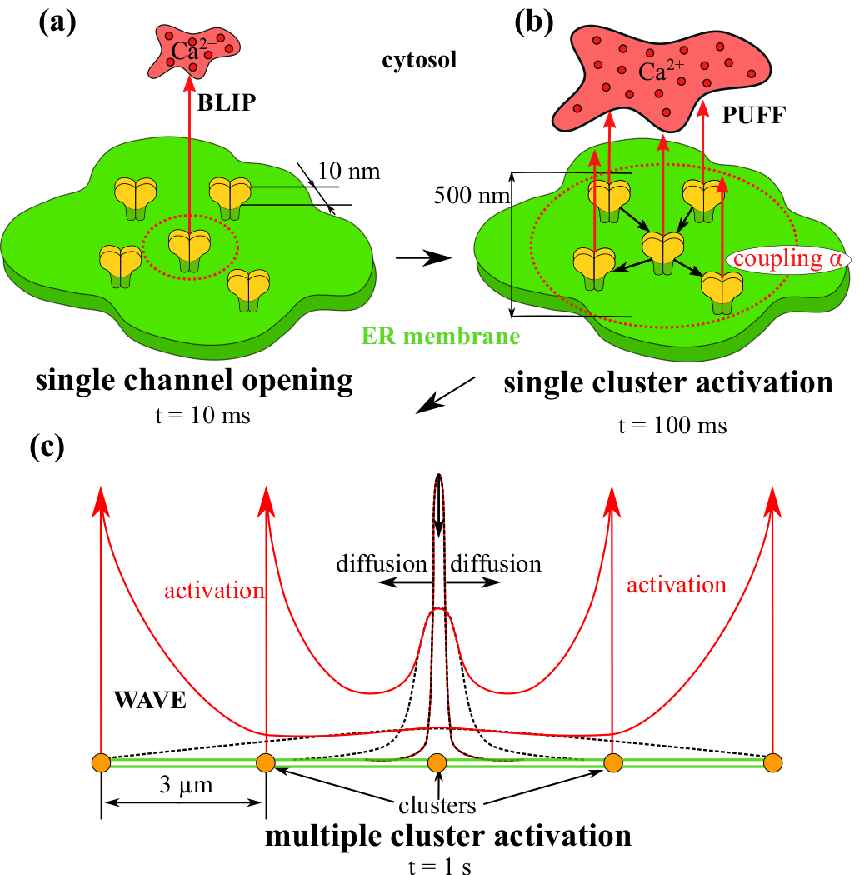}
	\caption{\label{fig:Conc} (Color online) Schematic representation of the spatial and temporal hierarchy of \ca release events. \textbf{(a)} A blip as the release from a single channel. The \textit{green plane} represents a part of the ER membrane, the \textit{red arrow} indicates \ca release from the opened channel indicated by the \textit{dotted circle}. \textbf{(b)} A puff is represented by multiple blips occurring in a single cluster of coupled channels. The coupling between channels (marked by \textit{black arrows}) is defined by the constant $\alpha$ in the nonlinear dependence (\ref{eq:cd}). \textbf{(c)} The wave propagation as a sequence of puffs (\textit{red lines}) caused by raised levels of [Ca\textupperscript{2+}]. In our modeling approach, we consider clusters of channels to be localized at single points in space. The activation of neighboring clusters is caused by Ca\textupperscript{2+} diffusion from the initial Ca\textupperscript{2+} release localized in the center (\textit{black dotted lines}).} 
\end{figure}

The assumption of fast equilibration to steady-state level $c_d$ in (\ref{eq:CaRel}) flattens out $c(t)$ from its blip amplitude $c_s$ of cytosolic \ca at a cluster. For $\lambda=10^3\,\text{s}^{-1}$ \cite{Rudiger2014}, it takes $\sim 1-10$ ms for \ca levels to reach cytosolic concentrations in the $10-10^3$ nM range. In what follows, we introduce a diffusion term to enable \ca to activate quiescent clusters by CICR thereby having blips combine to form puffs as shown in Fig.~\ref{fig:Conc}~(b). This is reminiscent of the fire-diffuse-fire model~\cite{Dawson1999}.

For a single cluster in a small domain of the membrane a computational study~\cite{Ruckl2015} using a hybrid reaction-diffusion model -- a Markov chain description as per DYK~\cite{DeYoung1992} and partial differential equations for diffusing \ca ions -- shows the dependence of the transition from puffs to waves on the numbers of channels in a cluster which initiate Ca\textupperscript{2+} release events under different IP\textunderscript{3} loads. The complexity of the analysis prompts us to look for a simplified model that could probe the temporal patterns observed and provide qualitative explanations for their origin via dynamical systems theory. We extend this study to a larger domain containing multiple clusters each described by (\ref{eq:active})--(\ref{eq:closed})~\cite{Rudiger2014}. While retaining the structure of the model given in \cite{Rudiger2014} we chose the parameters of \cite{Ruckl2015} in order to study the effect of different numbers of channels in each cluster and the interplay between them. This allows us not only to study the transition from puffs to waves but to obtain the characteristics of waves and oscillations that emerge. 


This is the justification of applying the principles of slow-fast reduction to the single channel and single cluster models -- we can elucidate how these reduced models may be coupled to account for physiological behavior that is manifest spatially.   

\subsection{\label{subsec:Model}Cluster-based spatial model}

In this section, we lay down our assumptions for building a spatially dependent model for a phenomenological representation of a Ca\textupperscript{2+} wave as a set of sequential Ca\textupperscript{2+} release from clusters [Fig.~\ref{fig:Conc}~(c)] using features of earlier models ~\cite{DeYoung1992,Li1994,Atri1993,Dawson1999,Rudiger2014,Ruckl2015}.  An examination of the dynamical regimes of individual cluster-level reductions that account for the blip to puff phenomena enables us to propose a model of diffusively coupled clusters in the ER membrane where Ca\textupperscript{2+} release occurs. We analyze the clusters separately using dynamical systems theory. We confirm the existence of two qualitatively different Ca\textupperscript{2+} behaviors in clusters with different numbers of channels, \emph{viz.}, excitable~(puff regime) and bistable~(wave regime)~\cite{Rudiger2014}. We shall probe whether introducing diffusive coupling and linear relaxation~(\ref{eq:CaRel}) (that can subsume effects including buffering~\cite{Sneyd1998} or other  modes of reducing spatial gradients) can facilitate the occurrence of oscillations in \ca release in the ER membrane.  This is incorporated into the model by modifying the equilibration eq.~(\ref{eq:CaRel}) by introducing a one-dimensional diffusive transport term
\begin{equation}
\frac{\partial c}{\partial t} = D\frac{\partial^{2}c}{\partial x^{2}} - \lambda\sum_i H(r_{cl}-|x-x_i|)\left(c-c_{d}(a_i)\right),\label{eq:CaSpace}
\end{equation}  
where $c_d(a_i)$ is given by eq.~(\ref{eq:cd}), $D$ is an effective diffusion constant, $x$ is a spatial coordinate, $H(x)$ is the Heaviside step-function ensuring that calcium releases into the cytoplasm   occur only at the clusters located at $x_i$, $i=\overline{1,L}$, and $r_{cl}$ is the cluster radius.
\begin{equation}
H(x) = \begin{cases}
1, & \text{if}\ x\ge 0, \\
0, & \text{otherwise}.
\end{cases}
\end{equation}
For the convenience of notation, we shall assume that the $i$ dependence on $a_i$ is implicit hereafter.

The model proposed in the current paper is of the fire-diffuse-fire type studied before~\cite{Dawson1999}. Unlike previous studies, we propose a direct link between the \ca level $c$ and the proportion of opened channels $a$ through nonlinear eq.~(\ref{eq:cd}). This also helps us to  associate cellular behavior of \ca releases on a spatial scale larger than the cluster size $r_{cl}$ to microscopic channel characteristics~\cite{DeYoung1992}.

System~(\ref{eq:active})--(\ref{eq:closed}) together with eq.~(\ref{eq:CaSpace}) is a reaction-diffusion system. The first term in the eq.~(\ref{eq:CaSpace}) corresponds to the smoothing of a Ca\textupperscript{2+} front and the spreading of [Ca\textupperscript{2+}] from the active clusters to the neighboring ones as sketched in Fig.~\ref{fig:Conc}~(c). In the second, reaction term, $c_d$ (\ref{eq:cd}) is a function of the number of opened channels and accounts for the opening of IP\textunderscript{3}R channels in the clusters.  The equilibration of cytosolic \ca to $c_d$ at rate $\lambda$ is a linear homeostatic reaction term that subsumes the action of pumps, leaks, and buffers, which drives the cytosolic [Ca\textupperscript{2+}] to the steady state value $c_d$.

We explore next how this model is able to generate waves and oscillations and characterize these phenomena in terms of the model parameters.

\section{RESULTS}

In subsection \ref{SC}, we apply the three-state model to a case of high [IP\textunderscript{3}] and corroborate the existence of a transition from excitable to bistable behavior in a cluster of channels~\cite{Rudiger2014}.  By continuously changing the rest \ca concentration parameter $c_0$ in a cluster with no opened channels,  we uncover the existence of two Hopf bifurcations. This illustrates the possibility that a single cluster could exhibit oscillations and channel activity and homeostatic mechanisms could regulate the state of the system to undergo dynamical regime shifts.  

Instead of letting $c_0$ be a fixed external control parameter that sets the level of \ca ions in the appropriate range for oscillatory behavior in the bifurcation analysis, we allow \ca levels to be dynamically set by diffusion, with the effective diffusion constant $D$ being the control parameter that sets the \ca levels locally and affecting the physiological output.  Diffusion levels gradients; in order to ascertain whether altering the levels between a cluster primed for bistability and that which is excitably monostable, we set up in subsection \ref{CB} a two-cluster model to study its phase diagram by bifurcation analysis.  Once again, we find the existence of two Hopf bifurcation points as $D$ is altered  and obtain the corresponding limit cycle trajectories. We associate those trajectories with the emergence of Ca\textupperscript{2+} oscillations.     

In the subsection \ref{CWO}, we apply our cluster-based model to study Ca\textupperscript{2+} release from a domain in the ER membrane with clusters containing different numbers of channels. We demonstrate how the interplay between excitable and bistable clusters is a mechanism for the emergence of Ca\textupperscript{2+} oscillations within the ER membrane. We also show that the cluster-based model is capable of exhibiting Ca\textupperscript{2+} waves that propagate throughout the membrane.

In the subsection \ref{ip3}, we show the role of the \ip unbinding in the termination of the \ca release from a bistable cluster. We account for the transitions between the upper and lower planes of the DYK cube for varying [IP\textunderscript{3}] levels.

\subsection{\label{SC}Ca\textupperscript{2+} dynamics at the scale of a single cluster} 

We study the three-state model in eqs.~(\ref{eq:active}), (\ref{eq:closed}) with parameters as shown in the last column of Table~\ref{tab:RudOrig}. We shall treat as control parameters for our bifurcation analysis $N$, the number of channels in a cluster and $c_0$, the rest concentration of \ca  ions when none of the channels are opened, with $(0.025\leq c_0\leq 0.6)\,\mu$M.  The choice of transition rates $k_i^+=0.02\,(\mu \text{M}\cdot\text{s})^{-1}$ and $k_i^-=1.56\,\text{s}^{-1}$ were determined by patch-clamp data on the probability distribution for channel opening typical for \textit{Xenopus} oocytes at high [IP\textunderscript{3}]~\cite{Ruckl2015}.   This has implications on the number of channels to be opened within each cluster in shaping the model building and results below.

\begin{figure}[h!]
	\includegraphics{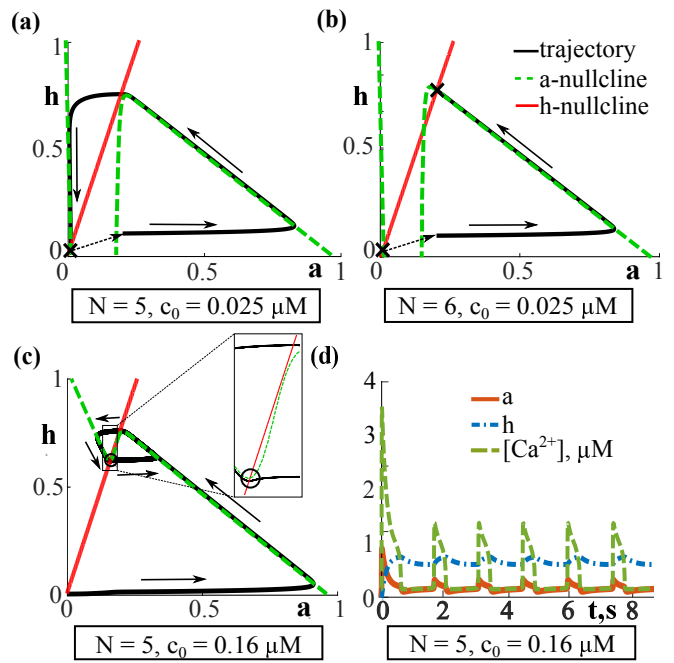}
	\caption{\label{fig:Osc} (Color online) The different regimes of Ca\textupperscript{2+} release in the two-dimensional three-state model (\ref{eq:active})--(\ref{eq:closed}) assuming $c=c_d$ given by (\ref{eq:cd})~\cite{Rudiger2014}. \textbf{(a)} The phase diagram of the excitable cluster ($N=5$), the \textit{red solid line} is $h$-nullcline for inhibition, the \textit{green dashed line} is $a$-nullcline for opened channel with at least three active  \ca bound, the \textit{black solid line} is a trajectory started from $a=0.2$, $h=0.1$ (\textit{black arrows}) and finished at the stable fixed point (\textit{black cross}). \textbf{(b)} The phase diagram of the bistable cluster ($N=6$), the diagram depicts two stable fixed points. The trajectory starts at $a=0.2$, $h=0.1$ and finishes at the upper fixed point. \textbf{(c)} Oscillations in the three-state model observed for $N=5$ and $c_0=0.16\,\mu \text{M}$. The only unstable fixed point is marked as the \textit{open circle} in the zoomed figure. \textbf{(d)} The temporal behavior of oscillations for a fraction of channels in open state denoted $a$ \textit{(red solid)}, a fraction of closed channels denoted $h$ \textit{(blue dot-dashed)}, and cytosolic [Ca\textupperscript{2+}] denoted $c$ \textit{(green dashed)}.}	
\end{figure}

We assume that cytosolic [Ca\textupperscript{2+}] equilibrates to the steady-state value $c_d$ very fast ($\lambda = 10^3\,\text{s}^{-1}$) in eq.~(\ref{eq:CaRel}). Therefore, we are able to represent the model of a single cluster dynamics by eqs.~(\ref{eq:active}), (\ref{eq:closed}) whose solutions are numerically obtained by standard Runge-Kutta methods and shown in the ($a$, $h$) phase plane in Figs.~\ref{fig:Osc}~(a),~(b), and~(c).  In the next subsections, this value will be reduced and the corresponding fast-slow decoupling will no longer be valid.

Figs.~\ref{fig:Osc}~(a) and (b) show system behavior as determined by the orientation and position of the $\dot{a}=f(a,h)=0$ (\textit{green dashed line}) and $\dot{h}=g(a,h)=0$ (\textit{red solid line}) nullclines and we  mark by `X'  the locations of the stable fixed points where the nullclines intersect.    The \textit{broken arrow} represents a perturbation of the system away from the stable fixed point closest to the origin.  After excitation, the system evolves following the trajectory which is guided by the  nullclines (\textit{green dashed} and \textit{red solid lines}). If the initial condition for the dynamical variables in Fig.~\ref{fig:Osc}~(a) lies to the right of the middle \textit{green dashed line}, the state variables return to the (only) fixed point following the extended trajectory in the phase plane shown in \textit{black}. In Fig.~\ref{fig:Osc}~(a) the trajectory comes back to the only fixed point and demonstrates that the monostable state is an excitable one, while in Fig.~\ref{fig:Osc}~(b) the system settles into the fixed point corresponding to the higher values of $(a,h)$, a bistable state.   The excitable dynamics creates a short-lasting puff when the Ca{\textupperscript{2+}} concentration is pulse-like. In contrast, the bistable cluster dynamics ensures that the  Ca{\textupperscript{2+}} concentration does not return to a base level but stays elevated for a longer period of time, the behavior associated with a long-lasting puff. The termination of such a long-lasting puff is attributed to IP\textunderscript{3} unbinding which is discussed in subsection~\ref{ip3}. Also, we shall show later the possible role of the intra-cluster, inter-channel coupling represented by $\alpha$ in eq. (\ref{eq:cd}) in accounting for termination of puffs.

We confirm the transition from monostable excitable and bistable behavior~\cite{Rudiger2014} in Figs.~\ref{fig:Osc}~(a) and (b). However, our use of parameters adapted from \cite{Ruckl2015} reveals the bistable state occurs when the number of channels is $N\geq 6$ unlike the higher value $N\geq 9$ in ~\cite{Rudiger2014}.  $N\sim 5,6$ was an average number of open channels when calcium waves were triggered as a function of increasing \ip the hybrid reaction-diffusion simulation~\cite{Ruckl2015}.  Having confirmed that changing the total number of  channels  that may be opened alters the dynamics, we then analyze the stability of the lower fixed point in the phase plane upon changes of parameter $c_0$.  As seen in Figs.~\ref{fig:Osc}~(a) and (b), the rest concentration of \ca ions is taken as $c_0=0.025\,\mu \text{M}$.  We change $c_0$ in a continuous manner $(0.025\leq c_0 \leq 0.6)\,\mu$M to probe qualitative shifts in system  dynamics.   The bifurcation diagram is shown in Fig.~\ref{fig:BifDiag}~(a) and the real and imaginary parts of the eigenvalues $(\lambda_{1,2})$ of the Jacobian of the system in Figs.~\ref{fig:BifDiag}~(c) and (d).  We find two Hopf bifurcation points ($\Re(\lambda_1$, $\lambda_2)=0$ and $\Im(\lambda_1$, $\lambda_2)\neq 0$) in the range of $c_0$ that  determine the onset or disappearance of limit cycles.

$c_0$ is an external control parameter in the bifurcation analysis with values in a physiologically plausible range.  It is the cytosolic [\ca] at a cluster before a Ca\textupperscript{2+} release event commences when no channels are opened.  The oscillatory behavior of the dynamical variables of the cluster with initial condition  $c_0=0.16\,\mu \text{M}$ for the concentration of Ca\textupperscript{2+} is shown in Figs.~\ref{fig:Osc}~(c) and (d) in two different representations.

\begin{figure}[h!]
	\includegraphics{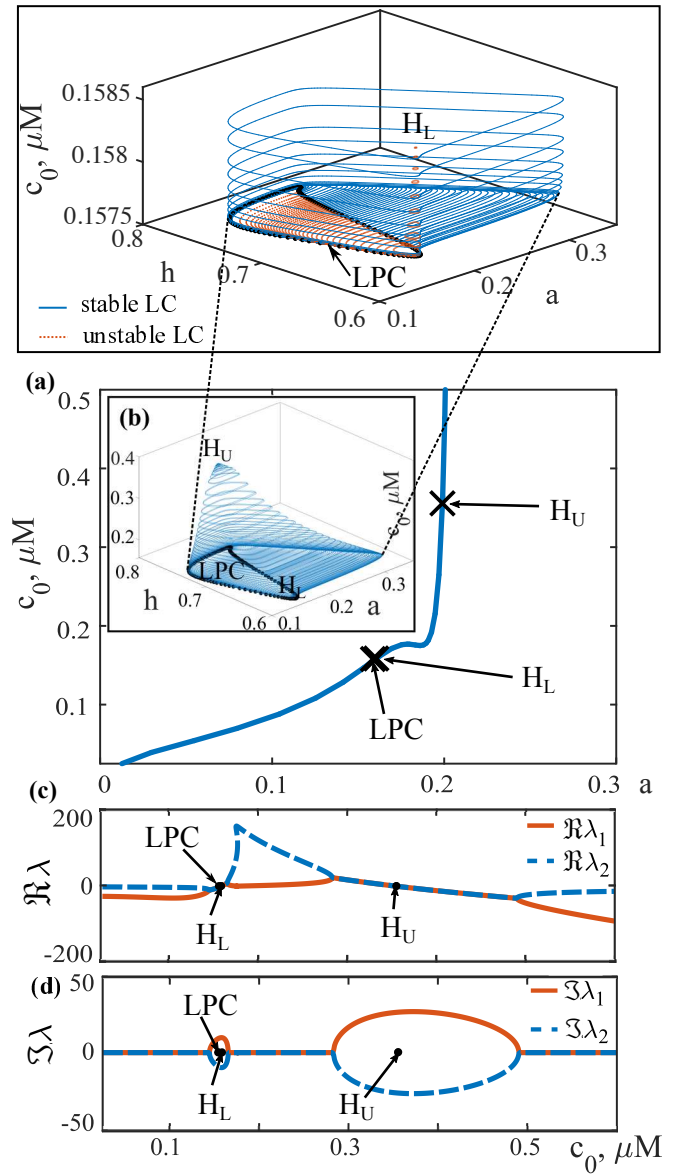}
	\caption{\label{fig:BifDiag} (Color online) \textbf{(a)} Bifurcation diagram of the two-dimensional system at $N=5$ plotted versus control parameter $c_0$, \textit{black crosses} labeled by H$_\text{L}$ and H$_\text{U}$ correspond to Hopf bifurcation points ($\Re\lambda=0$, $\Im\lambda\neq 0$), LPC corresponds to the value of $c_0$ where the fold bifurcation of two limit cycles occurs. \textbf{(b)} The stable limit cycle emerges at a supercritical Hopf point labeled as H$_\text{U}$. The continuation of the stable limit cycle shown by \textit{blue solid lines}. With the decrease in $c_0$ the unstable limit cycle occurs from subcritical H$_\text{L}$ point as indicated by \textit{red dashed lines} of small amplitude in zoomed view. \textit{Black dotted} }	
	\flushright \emph{(continued)}	
	\vspace{0.3cm}
	\hrule
\end{figure}

\begin{figure}
	\ContinuedFloat
	\caption{\textit{line} corresponds to the fold (limit point) bifurcation of two limit cycles marked as LPC. \textbf{(c)} The real parts of the first two eigenvalues of the system, which are zeros at Hopf points. \textbf{(d)} The imaginary parts of the eigenvalues of the system are non-zeros at Hopf points. Two saddle-node bifurcation points are also present in the diagram at $c_0\approx 0.18\,\mu$M, however, we do not show them in the plot as these points are not relevant to the analysis below and will no longer be discussed.}
\end{figure}

To study the stability of the limit cycles emerging from the Hopf bifurcation we perform a detailed dynamical systems analysis, presenting the results of the continuation and bifurcation analysis performed using the MatCont~\cite{Dhooge2008} package for Matlab in Fig.~\ref{fig:BifDiag}. This analysis keeps track of how a continuous change in parameters such as $c_0$ alters the fixed point shown in Fig.~\ref{fig:Osc}~(a) and, thus, the eigenvalues of the Jacobian around the altered fixed point change as well (see, e.g., Fig.~\ref{fig:Osc}~(c)).  In particular, we calculate the first Lyapunov coefficient $l_1$ (see~\cite{Kuznetsov1998}) derived from the normal form of the system of eqs.~(\ref{eq:active}), (\ref{eq:closed}), at each Hopf point (denoted H$_U$ and H$_L$), assuming $c=c_d$ as in eq.~(\ref{eq:cd}). If $l_1$ is positive (negative) the Hopf bifurcation is subcritical (supercritical) and the limit cycle is unstable (stable). In our case, $l_1=-2.9\times10^2$ at H$_\text{U}$, the upper  point in Fig.~\ref{fig:BifDiag} (b) (inset) where a stable limit cycle emerges via a supercritical Hopf bifurcation and $l_1=5.32\times10^4$ at the lower point H$_\text{L}$.  We start the continuation analysis from a maximum value of $c_{0}=0.5\,\mu$M where the fixed point is stable [top of the \textit{blue curve} in Fig.~\ref{fig:BifDiag}~(a)] as shown by the negative real and zero imaginary parts of the eigenvalues of Jacobian [right hand sides of Figs.~\ref{fig:BifDiag}~(c) and (d)]. Upon decreasing $c_0$ we reach the H$_\text{U}$ ($c_0 = 0.34\,\mu$M) point where the stable limit cycle emerges; we observe limit cycles with larger amplitudes with the further decrease in $c_{0}$ (\textit{blue cyclic trajectories} in Fig~\ref{fig:BifDiag}~(b)). The cyclic trajectory for the intermediate value of $c_{0}=0.16\,\mu$M is shown in Fig.~\ref{fig:Osc}~(c). After reaching H$_{\text{L}}$ point at $c_{0}=0.1585\,\mu$M the unstable limit cycle occurs as marked by \textit{red dashed cycles} of small amplitude in zoom of Fig.~\ref{fig:BifDiag}~(b). The stable cycle emerging from H$_\text{U}$ collides with the unstable one emerging from H$_\text{L}$ close to $c_0=0.158\,\mu$M.  This fold bifurcation of limit cycles is labeled by LPC (marked by \textit{black dotted trajectory}). The limit cycle solution ceases to exist while decreasing $c_0$ value lower then LPC point. Thus, we conclude that the stable limit cycle solutions exist in a small range of [Ca\textupperscript{2+}] concentration between H$_\text{U}$ and LPC points.

The bifurcation diagrams obtained from DYK~\cite{DeYoung1992} and Li-Rinzel~\cite{Li1994} models are different from the one presented above, owing to different assumptions about the speed of dynamical variables and different reduction schemes employed. In~\cite{DeYoung1992} and \cite{Li1994} [Ca\textupperscript{2+}] and [IP\textunderscript{3}] are dynamical, whereas we keep [IP\textunderscript{3}] at a high constant level. Moreover, the earlier models assume a well-mixed membrane and do not account for the clustering of channels that was introduced in \cite{Rudiger2014} setting the inactivation rate to be frozen at $\tilde{k}_i^+=k_i^+c_s$. We study the different Ca\textupperscript{2+} dynamics associated with IP\textunderscript{3}R clusters and physiological parameters for which the transitions for different kinds of behavior may occur.

The phenomena described above shows the Ca\textupperscript{2+} oscillations in a cytosolic [Ca\textupperscript{2+}] at a single cluster. The amplitudes of oscillations lie within a physiologically plausible range and their periods are similar to the oscillations found in several cells types~\cite{Callamaras1998}. However, various studies \cite{Skupin2008,Thurley2011} imply that \ca oscillations are collective in nature, and emerge when multiple clusters are involved.  In our model, we, therefore, cannot allow individual clusters to oscillate autonomously.  We shall show how it is only via the coupling between clusters that oscillations emerge, and switching off the inter-cluster coupling leads to the loss of rhythmic behavior. Thus, we next study the behavior of Ca\textupperscript{2+} release from arrays of clusters. 

\subsection{\label{CB}Cluster-based model of a discrete chain of clusters}

 Here we study the dynamics of a chain of clusters coupled by diffusing \ca ions. Firstly, we start by analyzing a simple example to illustrate the effect of coupling between neighboring clusters. Consider $L$ clusters coupled with the diffusion term in eq.~(\ref{eq:CaSpace}). The discrete form of eq.~(\ref{eq:CaSpace}) for the $i^{th}$ cluster is expressed in terms of the local \ca concentration $c_i$ whose rate of change is proportional to the spatial gradient of the diffusion current $J_i$ for $i=1,\ldots, L$.  The continuum diffusion equation is derived from 
\begin{eqnarray}
\frac{\partial c_i}{\partial t} =  \frac{1}{\ell} (J_i - J_{i-1}) &=& \frac{1}{\ell}\Big[D\left(\frac{c_{i-1}-c_i}{\ell}\right) - \\ \nonumber
&&- D\left(\frac{c_i - c_{i+1}}{\ell}\right)\Big]. \label{eq:disc}
\end{eqnarray} 
In our model with a discrete distribution of clusters, we shall take $\ell \approx 1.4\,\mu$m as the optimal distance between clusters adapted for the current model and $D$, the effective diffusion constant for \ca ions is taken to be $D=30\ \mu\text{m}^2/\text{s}^{-1}$ \cite{Allbritton1992}. Here we assume the clusters to be arranged in a ring topology, $c_{i+L}=c_i$.  We numerically solve the system (\ref{eq:active}), (\ref{eq:closed}) and (\ref{eq:CaSpace}) assuming the diffusion term as in eq.~(\ref{eq:disc}) and model parameters as in the previous section. Only the equilibration rate is modified for these simulations and set to $\lambda=230\,\text{s}^{-1}$, instead of $\lambda=10^3\,\text{s}^{-1}$ that was used earlier.  

\begin{figure*}[t]
	\includegraphics{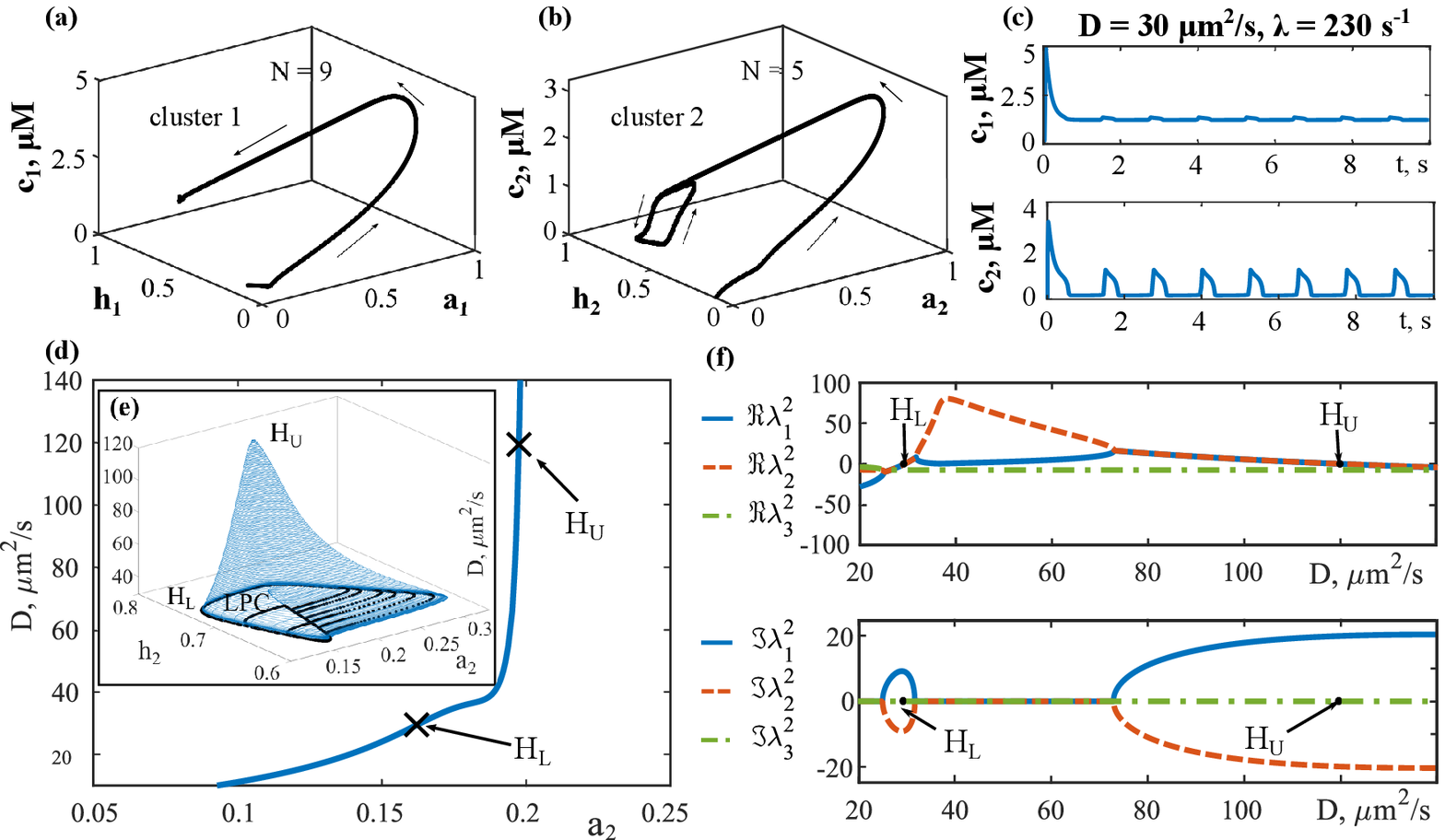}
	\caption{\label{fig:disc} (Color online) The initiation of Ca\textupperscript{2+} oscillations in the array of two clusters $D=30\,\mu \text{m}^2/\text{s}$, $\lambda=230\,\text{s}^{-1}$. \textbf{(a)} The trajectory in cluster $1$ with $N=9$ channels. \textbf{(b)} The oscillatory trajectory in cluster $2$ ($N=5$) triggered by the diffusion from the first cluster. \textbf{(c)} Oscillating Ca\textupperscript{2+} profiles in both clusters. \textbf{(d)} Bifurcation diagram of the system of two coupled clusters plotted versus diffusion coefficient $D$. The points H$_\text{L}$ and H$_\text{U}$ represent Hopf bifurcation points ($\Re\lambda=0$, $\Im\lambda\neq 0$). \textbf{(e)} The continuation of a limit cycle (blue lines) performed by analogy to Fig.~\ref{fig:BifDiag}~(b) varying $D$. Similarly, the black markers correspond to the fold bifurcation of limit cycles (LPC). \textbf{(f)} The real (\textit{top}) and imaginary (\textit{bottom}) parts of the eigenvalues of the system.}	
\end{figure*}

Here we consider the discrete lattice model for $L=2$, a typical unit cell of a discrete chain of clusters. The model parameters for both clusters are the same apart from the activatable numbers of channels in clusters, which are $N=9$ ($N=5$) in the first (second) cluster. In this case, our model is given by
\begin{eqnarray}
\frac{\dif a_{i}}{\dif t} &=& f(a_{i},h_{i},c_{i}), \; i=1,2 \label{eq:a_disc} \\ 
\frac{\dif h_{i}}{\dif t} &=& g(a_{i},h_{i},c_{i}), \; i=1,2 \label{eq:h_disc} \\ 
\frac{\dif c_{1}}{\dif t} &=& -\lambda [c_{1}-c_d(a_{1})] -\frac{2D}{\ell^2} (c_1 - c_2),\label{eq:c_disc1} \\ 
\frac{\dif c_{2}}{\dif t} &=& -\lambda [c_2 - c_d(a_{2}) ] - \frac{2D}{\ell^2} (c_2 - c_1).\label{eq:c_disc2}
\end{eqnarray}
We present the results of this unit cell $L=2$ model in Fig.~\ref{fig:disc}~(a), (b) and (c). Note, that the unit cell case is a particular example of a chain of clusters of length $L$. We use the discrete chain approximation to illustrate the behavior of the clusters in a ``one-dimensional membrane''. 

Unlike the single-cluster approach, calcium concentration $c$ in the cytosol is explicitly added to the model, the equilibration rate $\lambda$ is reduced and the diffusion term is introduced in eqs.~(\ref{eq:c_disc1}), (\ref{eq:c_disc2}).  The assumption that $c$ will be ``slaved'' to $c_d$ is inapplicable; we need to account for a separate dynamics of \ca cytosolic concentration away from the steady-state value $c_d$.

The oscillatory trajectories are exhibited in clusters~$1$ [Fig.~\ref{fig:disc}~(a)] and~$2$ [Fig.~\ref{fig:disc}~(b)] as the consequence of a diffusive interaction between the clusters. Notably, the concentration of Ca\textupperscript{2+} at cluster~$1$ exhibits small amplitude oscillations [Fig.~\ref{fig:disc}~(c) \textit{top}] around the fixed point with high $c_1$ [here the fixed point is associated with the higher fixed point in Fig.~\ref{fig:Osc}~(b) for the single cluster case but now we solve the~$6$--variable system]. The influx from the cluster~$1$ causes the emergence of Ca\textupperscript{2+} oscillations in cluster~$2$~[Fig.~\ref{fig:disc}~(c) \textit{bottom}]. In order to study this effect, we perform the continuation analysis of the system of two coupled clusters given by eqs.~(\ref{eq:a_disc})--(\ref{eq:c_disc2}) similarly to a single cluster approach described earlier. The dimensionality of the system used in the bifurcation analysis remains~$6$, but we represent the results as two (three) dimensional plots to depict the governing dynamics at each cluster. We plot the bifurcation diagram shown in Fig.~\ref{fig:disc}~(d) which represents the dependence of the $a_2$ value of the fixed point versus $D$ as the control parameter. The effect of Ca\textupperscript{2+} flux from cluster~$1$ to cluster~$2$ is proportional to $D$ which controls the strength of coupling. We observe a similar diagram as shown in Fig.~\ref{fig:BifDiag}. We also observe H$_\text{U}$ and H$_\text{L}$ points where stable and unstable limit cycles occur, respectively. From the eigenvalues plotted in Fig.~\ref{fig:disc}~(f), we confirm that H$_\text{U}$ and H$_\text{L}$ are Hopf bifurcation points. The continuation of the stable limit cycle occurring at H$_\text{U}$ [Fig.~\ref{fig:disc}~(e)] shows that the periodic solutions in the cluster~$2$ exist in the range $(29\leq D \leq  120)\,\mu \text{m}^2/\text{s}$.  This brings the range of concentrations of \ca to the one shown in Fig.~\ref{fig:BifDiag} for the same $\lambda$. Thus, we conclude that the emergence of oscillations in the second cluster depends on the concentration influx from cluster~$1$ to cluster~$2$ which raises the level of concentration in cluster~$2$ to a range that drives oscillations in the single cluster approach.

Having accounted for the ability of diffusive couplings to bring about oscillations in a coupled cluster system when the individual clusters exhibit non-oscillatory dynamics on their own, we extend the model to a continuum description.  In the next subsection, we apply this mechanism to a domain of the membrane in order to qualitatively demonstrate the emergence of Ca\textupperscript{2+} waves and oscillations in the ER membrane.

\subsection{\label{CWO}Calcium Waves and Oscillations in the cluster-based model}

As shown above, the behavior of the three-state model for a single cluster  depends on the number of channels $N$ in a cluster that are \ip bound and may be activated or deactivated by Ca\textupperscript{2+}, namely, excitable for small $N$ transitioning for large $N$ to bistable dynamics. While the ER membrane would typically contain clusters of different numbers of channels, we study the simplest heterogeneous scenario with one bistable cluster with $N=9$ and the multiple excitable clusters with $N=5$. Here and furthermore we refer to a $N=9$ cluster as bistable and a $N=5$ cluster as excitable as per their behavior in the isolated single cluster case, even though it is the entire system whose stability matters. The choice of this figures is based on the study~\cite{Ruckl2015}, where the average number of activatable channels in \textit{Xenopus} oocytes ER membranes tend to $N=5$ with the increase in [IP\textunderscript{3}], the distribution of numbers of channels appears to be very close to a Poisson distribution~\cite{Ruckl2015,Taufiq-Ur-Rahman2009}. We chose this numbers to be consistent with~\cite{Ruckl2015} and propose the modeling of more complex heterogeneous systems for future study. In the chosen situation the excitable behavior may be associated with short-lasting puffs while the bistable behavior may be associated with waves or long-lasting puffs, where the exit from the long-lasting puffs that is thought to be driven by dissociation of \ip or other mechanisms which are discussed in subsection~\ref{ip3}. The exit from the long elevated high \ca concentrations might be performed through the other mechanisms such as change in coupling strength $\alpha$, maximal \ca elevation $c_s$ or transition rate $k_i^+$.  
 
In this section, we model a domain of an ER membrane which is assumed to contain multiple excitable clusters ($N=5$) and the only one bistable ($N=9$) cluster in order to study the interplay between excitable and bistable clusters.  We demonstrate the emergence of oscillations  and quantify the characteristics of the phenomena such as front velocity and period.

\begin{figure}[h]
	\includegraphics{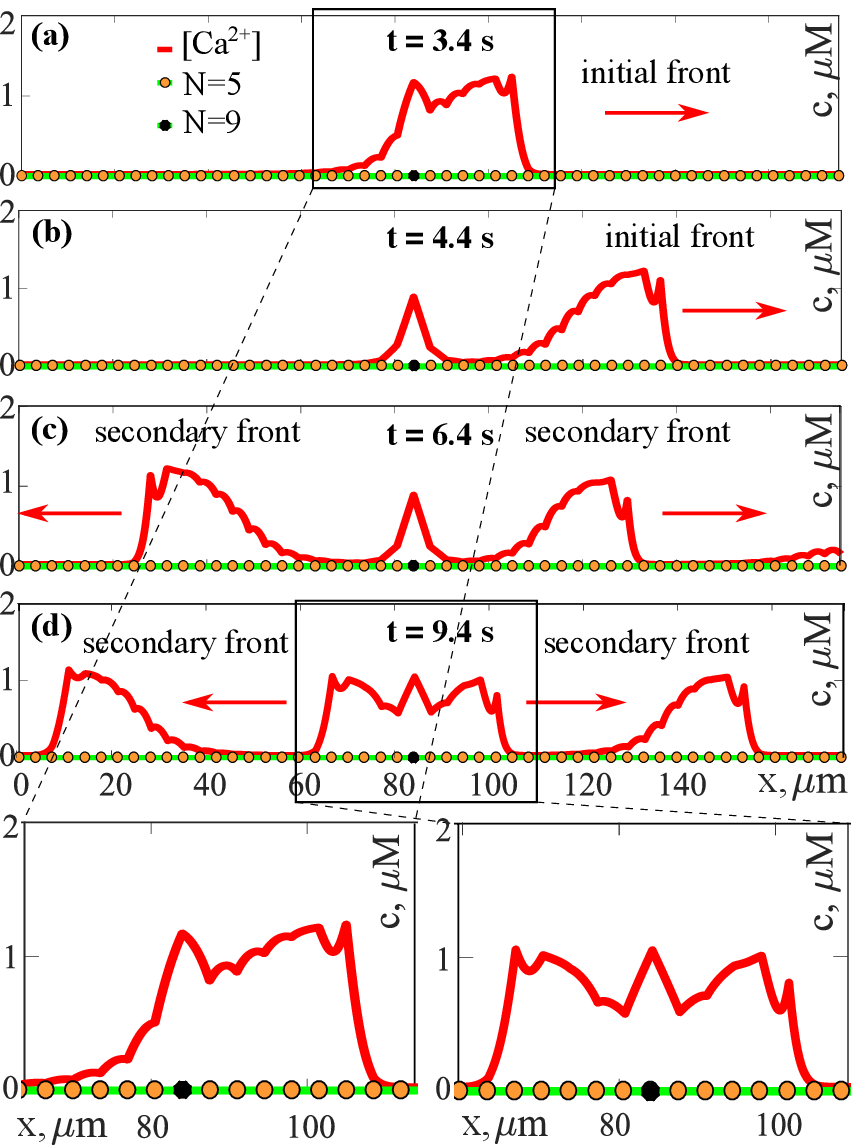}
	\caption{\label{fig:Oscillations} (Color online) Ca\textupperscript{2+} front propagation from the spatial model for two types of clusters ($D=30\,\mu \text{m}^2/$s, $\lambda=230\,\text{s}^{-1}$). The \textit{orange dots} represent excitable clusters with $N=5$ channels as long as the \textit{large black dot} corresponds to a bistable cluster with $N=9$ channels. \textbf{(a)} The propagation of a front caused by the raised initial [Ca\textupperscript{2+}] in the left part of the membrane. \textbf{(b)} The Ca\textupperscript{2+} peak occurs at the bistable cluster. \textbf{(c)} The diffusion from the peak causes initiation of a secondary wave. \textbf{(d)} The front propagates after that subsequent front occurs. The waves occur periodically from the bistable cluster.}  
\end{figure}

We numerically solve the eqs.~(\ref{eq:active}), (\ref{eq:closed}) and (\ref{eq:CaSpace}) in the chosen one-dimensional domain with open boundaries and Neumann boundary conditions $c^{'}_{x}(x=0,L) = 0 $. Let us assume the part of ER membrane with the initial distribution of cytosolic [Ca\textupperscript{2+}] $c(x,0)=c_0 + A\exp\{-[x^2/(2\sigma^2)]\}$, where $A=2\,\mu$M, $\sigma=0.1\,\mu\text{m}$, which travels from the left part of a membrane as shown in Fig.~\ref{fig:Oscillations}~(a).  Calcium release upon activation occurs only at channel clusters $i$ in eq.~(\ref{eq:CaSpace}) via $c_d(a)$ and cytosolic concentration of \ca is raised above basal level $c_0$ by diffusion between clusters a constant distance $\ell=3.5\,\mu \text{m}$ apart. The results depicted in Fig.~\ref{fig:Oscillations} are obtained for equilibration rate $\lambda=230\,s^{-1}$. The effect of changing the values of $D$, $\lambda$, and $\ell$ upon qualitatively differing \ca release response is discussed further in this section. 

In Figs.~\ref{fig:Oscillations}~(c) and (d) we observe the emergence of Ca\textupperscript{2+} oscillations at the bistable cluster (\textit{black dot}) triggered by the initial front [Fig.~\ref{fig:Oscillations}~(a)]. The front occurs from an initial Ca\textupperscript{2+} increase close to the point $x=0\,\mu \text{m}$ and diffuses throughout the membrane. While the initiation event is set by choosing initial condition in our deterministic model it reflects the potential occurrence of an initial peak by either an internal or external stimulus or a stochastic fluctuation. After passing the bistable cluster the initial front raises the concentration at this cluster to the value of $c$ which corresponds to the upper fixed point in Fig.~\ref{fig:disc}~(a). This behavior induces a long-lasting state which is terminated by IP\textunderscript{3} unbinding, a feature which we discuss later. Therefore, in Fig.~\ref{fig:Oscillations}~(b) a residual peak occurs at cluster $N=9$. Thereafter, two secondary fronts emerge from the residual peak as shown in Fig.~\ref{fig:Oscillations}~(c). As we consider deterministic oscillations this process repeats periodically with a constant period [Fig.~\ref{fig:Oscillations}~(d)].  

This behavior can be observed because of two different regimes underlying Ca\textupperscript{2+} release in clusters: excitable for short-lasting puffs and bistable for long-lasting events. Our model shows that bistable clusters can produce long-lasting puffs that might cause the emergence of Ca\textupperscript{2+} oscillations in cellular membranes.

The diffusion coefficient $D$ and relaxation parameter $\lambda$ incorporate spatial effects without explicitly modeling buffers in the membrane. As these are likely to influence the characteristics of the \ca waves and oscillations we vary $D$ to study how these characteristics change. Furthermore, by changing equilibration rate $\lambda$ we can model the inclusion of a linearized reaction Ca\textupperscript{2+} leak flux and pump action terms in the eq.~(\ref{eq:CaSpace}). There are different timescales for the activation and inhibition of the channels ``hidden'' in eqs.~(\ref{eq:active}), (\ref{eq:closed}). The action of pumps occurs on much slower timescales compared to $\lambda$ chosen. 

\begin{table}[h]
	\caption{\label{tab:SpatialModel}Parameters used in the cluster-based reaction-diffusion model.}
	\begin{ruledtabular}
		\begin{tabular}{c c c}
			Parameter & Value & Description \\ \hline	
			$c_0$ & $0.025\,\mu \text{M}$ & Rest level of [Ca\textupperscript{2+}] \\
			$N$ & $5$ or $9$ & Number of activatable channels \\
			& & in a cluster (IP\textunderscript{3} bound) \\ 
			$\lambda$ & $230\,\text{s}^{-1}$ & Equilibration rate \\
			$D$ & $30\,\mu \text{m}^2/$s & Diffusion coefficient \\ 
			$\ell$ & $3.5\,\mu \text{m}$ & Inter-cluster distance \\ 
			$r_{cl}$ & $0.1\,\mu \text{m}$ & Cluster radius
		\end{tabular}
	\end{ruledtabular}
\end{table}  

\begin{figure}[h!]
	\includegraphics{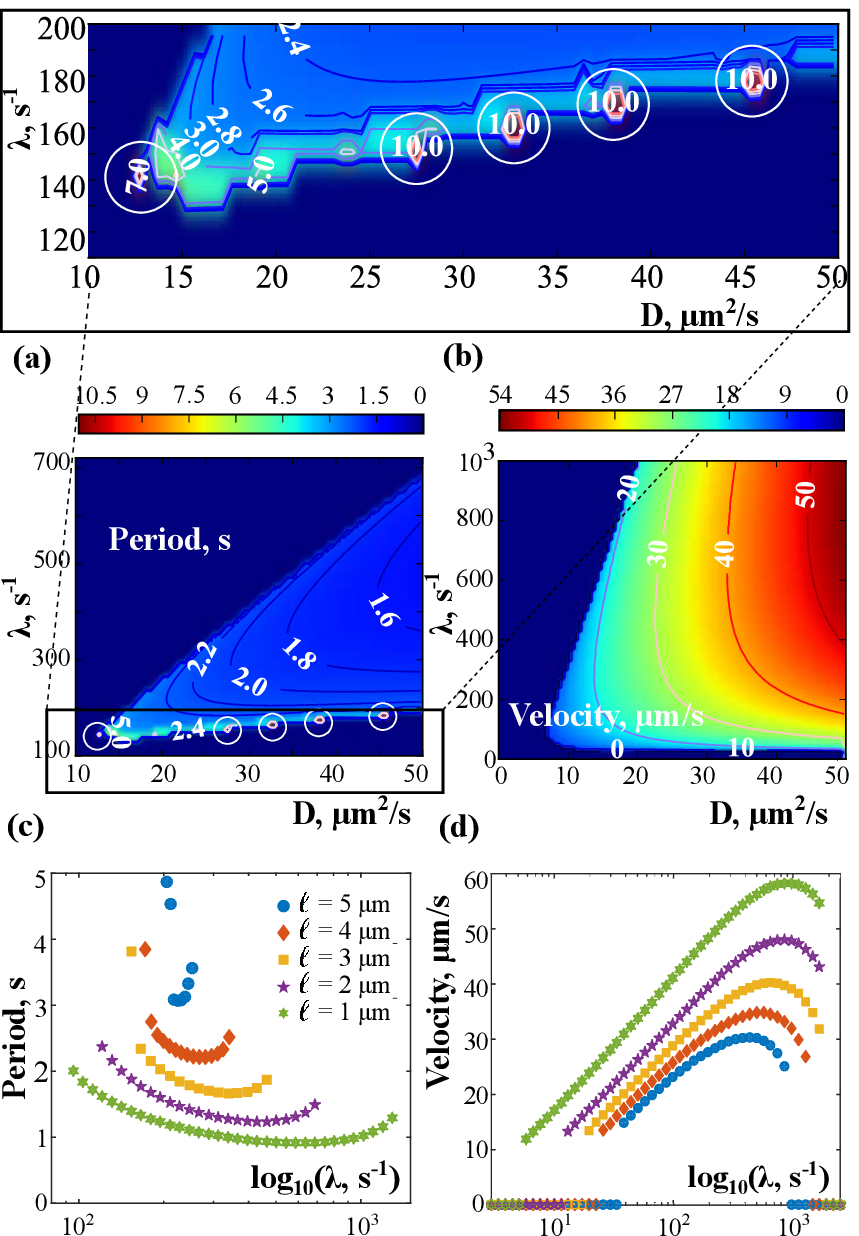}
	\caption{\label{fig:Map} (Color online) \textbf{(a)} The phase diagram which represents different kinds of \ca dynamics such as puffs, waves, and oscillations. The domain with zero periods [\textit{dark-blue zone} in (a)] contains both puffs and waves in a co-existing manner. The \textit{triangular light-blue domain} with non-zero periods contains oscillatory behavior of the system. The \textit{white circles} indicate the regions with \ca alternans. \textbf{(b)} The phase diagram representing two regimes of \ca dynamics such as puffs and waves in the case when all the clusters are excitable. Zero velocity domain in \textit{dark-blue} corresponds to puffs and abortive waves, while the non-zero-velocity domain (shown in \textit{blue} to \textit{red} colors) represents waves. \textbf{(c)} The dependence of the period on $\lambda$ for different inter-cluster distances, $D=30\, \mu \text{m}^2/$s. \textbf{(d)} The dependence of velocity on $\lambda$, the distances, and the diffusion coefficient are the same as in (c).}    
\end{figure} 

In Fig.~\ref{fig:Map}~(a) we show the values of the average time calculated between two subsequent Ca\textupperscript{2+} release events at each cluster. This value appears to be almost constant. We map the change in a period of \ca oscillations to physiological ranges of $D$ and $\lambda$ [Fig.~\ref{fig:Map}~(a)]. Dark regions correspond to a non-oscillatory response, while oscillations occur within the purple triangular zone.  A curious effect is observed at the lower edge of the triangle in the range $(100 \le \lambda \le 200)\,\text{s}^{-1}$. The light spots marked by white circles correspond to the region on the diagram where oscillations exhibit periods $\approx 10$ s and belong to \ca alternans -- a pulse travels alternately to the right, followed by another to the left and so on (Fig.~S2). Under given initial conditions, this effect can be explained by the fact that the concentration diffusing from the irregular cluster is sufficient to excite a neighboring cluster only from one side at a time. Thus, we observe the oscillations occurring on alternate sides of the bistable cluster sequentially. In Fig.~\ref{fig:Map}~(b) we show the velocity of propagation of the initial front for a similar range of $D$ and $\lambda$: the dark regions corresponding to abortive waves or puffs and the light trapezium corresponds to propagating Ca\textupperscript{2+} waves.  

The distance between clusters in the various cell types lies in range $1-7\,\mu \text{m}$~\cite{Rudiger2014a}. Figs.~\ref{fig:Map}~(c) and (d) show the dependence of the periods and the velocities on $\lambda$ for different values of the distance $\ell$ between the clusters. In both graphs, we consider the effective diffusion constant to be fixed at a typical value of $D=30\,\mu \text{m}^2/$s. The periods of the oscillations [Fig.~\ref{fig:Map}~(c)] generally reach larger values for long distances between clusters. This can be explained by the fact that the front travels a longer distance in order to reach neighboring clusters. Interestingly, the oscillatory behavior exists in wider ranges of $\lambda$ for smaller distances between the clusters [Fig.~\ref{fig:Map}~(c)]; oscillations are more robust to the equilibration process if the distances between clusters are smaller. The same processes define the velocities [Fig.~\ref{fig:Map}~(d)]; these are higher for shorter distances because a front reaches a neighboring cluster faster and is able to trigger a response there. The data with comparatively large periods $\approx 4-5\,\text{s}$ for $\ell=5\,\mu \text{m}$, $\ell=4\,\mu \text{m}$ and $\ell=3\,\mu \text{m}$ in Fig.~\ref{fig:Map}~(c) exhibits the non-regular alternans emerging similarly as in Fig.~\ref{fig:Map}~(a). 

\subsection{\label{ip3}IP\textunderscript{3} dependence}

\begin{figure*}
	\includegraphics{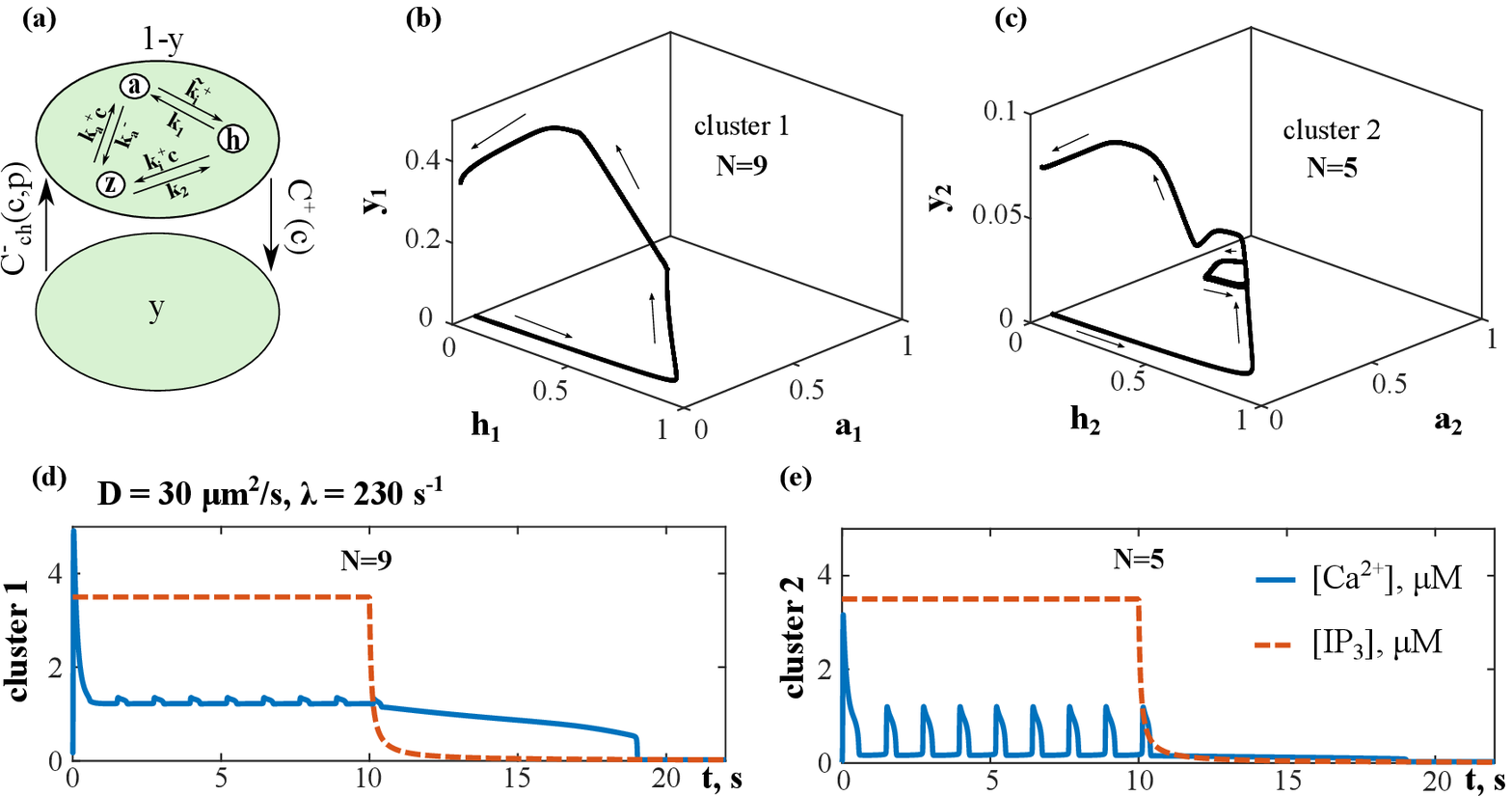}
	\caption{\label{fig:p_change} (Color online) The \ip dependent model for two coupled clusters. \textbf{(a)} The scheme of an extension of the three-state model introduced in Fig.~\ref{fig:DYKRud}~(c). The transitions between the upper and the lower plane of the DYK cube are governed by the steady-state rates $C^+$ and $C^-_\text{ch}$ in analogy to Li-Rinzel~\cite{Li1994}. \textbf{(b)}, \textbf{(c)} The trajectories in the $(a,h,y)$ phase space, where $y$ corresponds to the fraction of the channels in a cluster occupying the lower plane of the DYK cube, $N$ is the number of activatable channels in a cluster. \textbf{(d)}, \textbf{(e)} The oscillatory regime corresponding to Fig.~\ref{fig:disc} (\textit{blue solid lines}) for high [IP\textunderscript{3}] is terminated due to the decrease in the [IP\textunderscript{3}] (\textit{red dashed lines}).}	
\end{figure*}

The assumption of high levels of [IP\textunderscript{3}] leads to a sustained \ca release from the bistable cluster. In the current subsection, we aim to link the termination of this event to the change in the [IP\textunderscript{3}] levels [\textit{red dashed lines} in Figs.~\ref{fig:p_change}~(d),~(e)] on timescales slower than the oscillations found in previous subsections [\textit{blue solid lines} in Figs.~\ref{fig:p_change}~(d),~(e)]. 

The extension of the model is sketched in the Fig.~\ref{fig:p_change}~(a). In the case of two coupled clusters the \ip dependent model (\ref{eq:a_disc}), (\ref{eq:h_disc}) reads as
\begin{eqnarray}
\frac{\dif a_{i}}{\dif t} &=& f_p(a_{i},h_{i},y_{i},c_{i}), \; i=1,2 \label{eq:a_2ip3} \\ 
\frac{\dif h_{i}}{\dif t} &=& g_p(a_{i},h_{i},y_{i},c_{i}), \; i=1,2 \label{eq:h_2ip3} \\ 
\frac{\dif y_{i}}{\dif t} &=& C^+(c_{i})(1-y_{i})-C^-_\text{ch}(c_{i},p)y_{i},\label{eq:y_2ip3} 
\end{eqnarray}
where $f_p$ and $g_p$ are given in eqs.~(\ref{eq:a_z}) and (\ref{eq:h_z}), $C^+$ and $C^-_\text{ch}$ are given in eqs.~(\ref{eq:c_pm}) and (\ref{eq:c_m_ch}), respectively. The \ip changes as
\begin{equation}
p(t) = \begin{cases}
p_0, & \text{if}\ t\le t_\text{dur}, \\
\frac{p_a \sigma_0^2}{\sigma_0^2 + 2 D_p t} \operatorname{e}^\frac{-x^2-y^2}{4 D_p t + 2 \sigma_0^2}, & \text{otherwise},
\end{cases}
\end{equation}
where the initial value of [IP\textunderscript{3}] at a cluster $p_0=3.5\,\mu$M, the duration of the IP\textunderscript{3} stimulus is $t_{\text{dur}}=10\,$s. We assume the shape of the initial [IP\textunderscript{3}] distribution to be Gaussian of amplitude $p_a=4.6\,\mu$M and with variance $\sigma_0^2 = 1\,\mu \text{m}^2$ (see Supporting Material Sec.~S3). After stimulus terminates, [IP\textunderscript{3}] diffuses in infinite 2D domain ($D_p=10\,\mu\text{m}^2/$s~\cite{Dickinson2016}). We calculate time traces shown by \textit{red dashed lines} in Fig.~\ref{fig:p_change}~(d) and (e) at clusters positioned at $x = \pm (\ell + r_{cl})/2$ and $y=0$ (here the distance between clusters $\ell=1.4\,\mu$m, cluster radius $r_{cl}=0.1\,\mu$m). \ca dynamics remains unaltered as in eqs.~(\ref{eq:c_disc1}), (\ref{eq:c_disc2}).

The trajectories in $(a,h,y)$ phase space for both clusters are exhibited in Figs.~\ref{fig:p_change}~(c),~(d). The global oscillations occur in our model at [IP\textunderscript{3}]$\approx 3-4\,\mu$M, which is compatible with the levels considered in \cite{Marchant1999}. The termination of the oscillations [solid blue lines in Fig.~\ref{fig:p_change}~(e),~(d)] at both clusters is caused by the decrease in [IP\textunderscript{3}] under a level of several tens of nM, which is typically taken as the rest value of the [IP\textunderscript{3}]~\cite{Ruckl2015}. It might be useful to reconsider the influence of the calcium pumps in the termination process as we had argued that their effects are subsumed in a linearized equilibration rate $\lambda$.  We have checked that the effect of introducing an explicit pump term \cite{DeYoung1992} on the wave termination behavior can be accommodated by adjusting $\lambda$.

\section{\label{sec:Disc}DISCUSSION AND CONCLUSIONS}

In this paper, we presented a reaction-diffusion model of a membrane containing clusters of \ip receptor channels of the fire-diffuse-fire~\cite{Dawson1999} type but with a mechanistic description of channel firing similar to the spatial Fitzhugh-Nagumo model~\cite{FitzHugh1961,Jones1984}.  This mechanistic model was built on top of a dynamical systems analysis of a previously proposed simplification~\cite{Rudiger2014} of the DYK model~\cite{DeYoung1992}.  We noticed that incorporating the transition rates given in~\cite{Ruckl2015} into the model revealed the emergence of \ca oscillations even in a single cluster which inspired the membrane level formulation of a model of \ca puffs, waves, and oscillations whose characteristics were quantified under varying cluster configurations and physiological conditions. Our model allows us to observe \ca waves with the velocities in agreement with experiments by Marchant et al.~\cite{Marchant1999} for various distances between clusters and equilibration rates~[Figs.~\ref{fig:Map}~(b) and (d)].

Here we propose the mechanism of initiation of oscillatory behavior in the membrane that consists of the following sequence.

\begin{itemize}
	\item 
	Upon receiving an \ip stimulus the membrane configures \ip receptor channels to form clusters with different numbers of channels~\cite{Taufiq-Ur-Rahman2009} with bound \ip that are primed for activation by \ca binding. 
	\item
	If the number of primed channels in a cluster is higher than a threshold value, then the cluster is bistable. When an initial surge of \ca passes the bistable cluster [Fig.~\ref{fig:Oscillations}~(b)] it drives Ca\textupperscript{2+} concentration there into the higher fixed point of the dynamics as shown in Fig.~\ref{fig:disc}~(a). Sustained \ca elevation at a bistable cluster corresponds to a long-lasting release event.
	\item 
	The residual [Ca\textupperscript{2+}] at the bistable cluster diffuses to the neighboring clusters.
	\item
	This drives [Ca\textupperscript{2+}] at the neighboring clusters into an oscillatory regime as in Fig.~\ref{fig:disc}. These oscillations spread to further clusters due to diffusive coupling.
\end{itemize}
It is generally known that high \ca concentrations might lead to cell death and this renders the presence of bistability and consequent sustained \ca release in our model an unhappy feature. Due to fast luminal \ca refilling~\cite{Lopez2016} sustained \ca release, such as what follows from a transition to the upper fixed point of the bistable regime, cannot be terminated by ER depletion at the timescales considered by current paper.   The mechanistic origins of our model enable us to propose possible physiological possibilities that might explain the termination of \ca release from the bistable cluster.  

\begin{itemize}
\item \textit{\ip unbinding.} In the full DYK approach, R{\"{u}}diger et al.~\cite{Rudiger2012} attribute the termination of long-lasting \ca release events to \ip unbinding. A simplified model with inclusion of the lower plane of the DYK cube shown in subsection \ref{ip3} has also displayed the termination of a \ca wave in the bistable cluster in Fig.~\ref{fig:p_change}. Thus, we argue that neglecting \ip unbinding completely might lead to results which seem unrealistic. Therefore, the limits of the applicability of our model are bound to the change of the \ip levels in the cell. 

	
\item \textit{The uncoupling of \ipr channels.} Our bifurcation analysis shows that the system ceases to be bistable if the coupling between channels $\alpha$ decreases. Hence, if $\alpha$ decreases on timescales slower than oscillation periods observed in the model, the long-lasting event can terminate (see Fig.~\ref{fig:termination}). In this paper, we have considered $\alpha$ to be constant, which might not always be the case in real clusters. It is known from experiments that the coupling (uncoupling) of \ipr channels is caused by \ip binding (unbinding)~\cite{Tojyo2008,Taufiq-Ur-Rahman2009,Smith2009b}; the mechanisms which connect channels remain largely unknown, however. If an increase in $p$ leads to a fast (hundreds of ms according to~\cite{Smith2009b}) increase in coupling strength $\alpha$ could acquire a value as taken in this paper.    Following the initial stimulus that triggers this increase of \ip and its consequent effect on $\alpha$, a gradual diminution of \ip levels could still restrict the dynamical states to the upper place of the DYK cube, while reducing the coupling between channels in a cluster.  This  would lead to  the termination of puffs on the timescales defined by the change in [IP\textunderscript{3}] levels. However, only decreasing $\alpha$ under $0.1$ [Fig.~\ref{fig:termination}~(a)], which corresponds to a \ca concentration at a single cluster, leads to the exit from the bistable state. A mechanism which causes decoupling of channels in large clusters has been discovered in the RyR clusters~\cite{Marx2001} although a similar property has not been reported for \ip receptors.

\begin{figure}[h]
	\includegraphics{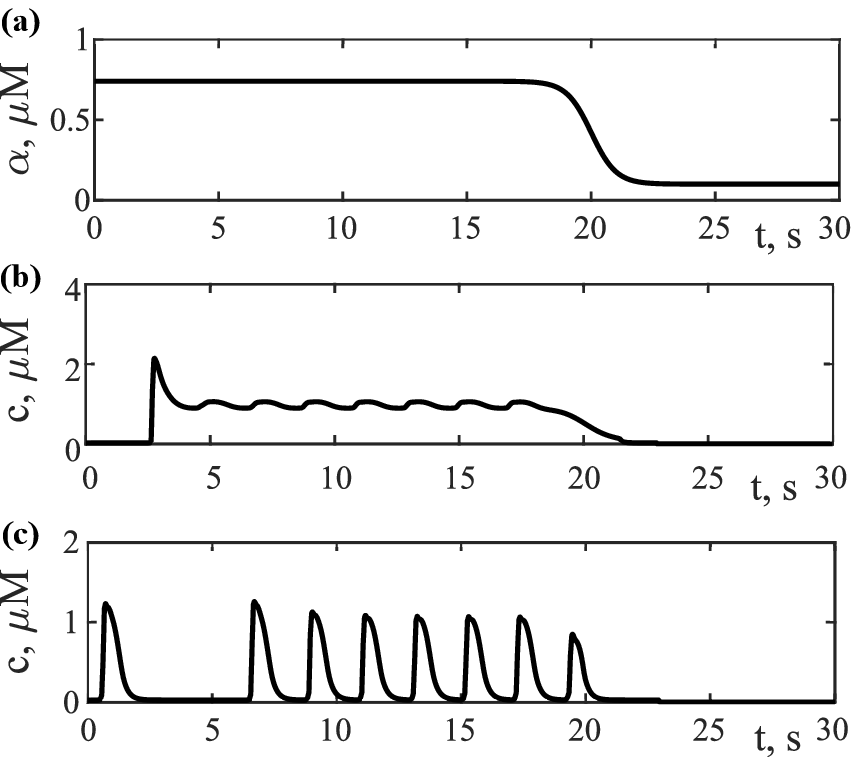}
	\caption{\label{fig:termination} \textbf{(a)} The change of the coupling strength $\alpha$ in eq.~(\ref{eq:cd}) on a slow timescale caused by the uncoupling of \ipr channels due to \ip unbinding. \textbf{(b)} The termination of the sustained \ca release at the bistable cluster shown in Fig.~\ref{fig:Oscillations}. \textbf{(c)} The termination of the oscillatory \ca release in one chosen excitable cluster shown in the Fig.~\ref{fig:Oscillations}.}    
\end{figure}

\item \textit{ER depletion.} In this study, we assume the same availability of \ca in the luminal pool in (\ref{eq:cd}). Even though the ER pool can be large and its fast refilling can prevent local depletion~\cite{Lopez2016}, other studies suggest that the large pool can be completely depleted by high-amplitude sustained \ca releases~\cite{Ullah2012}. 

\item \textit{Stochastic termination.} The bistability found in the single-cluster model can also be terminated stochastically. Fig.~\ref{fig:stochastic} shows the distribution of inter-puff intervals (IPIs) obtained from the Langevin model based on eqs.~(\ref{eq:active})--(\ref{eq:closed}). We consider the additive Wiener noise as shown in (\ref{eq:stoch_a}), (\ref{eq:stoch_h}) and compare the results with experimental puff data~\cite{Dickinson2012}. The duration of the events lies in a range of hundreds of ms which is also consistent with the experimental data available.

\begin{figure}[h]
	\includegraphics{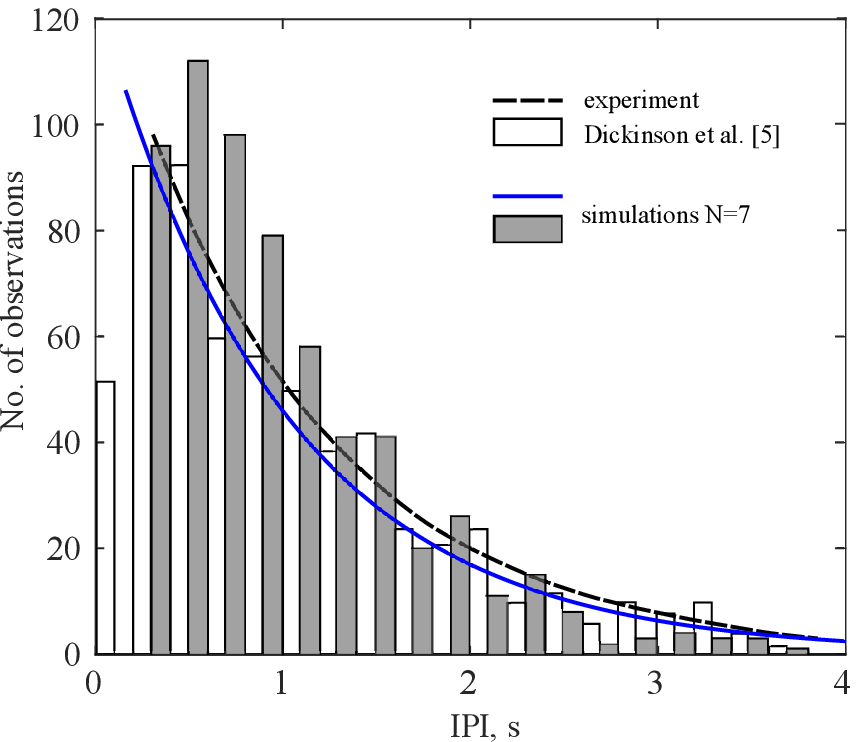}
	\caption{\label{fig:stochastic} (Color online) Inter-puff intervals of the stochastic single-cluster model (\ref{eq:stoch_a}), (\ref{eq:stoch_h}), (\ref{eq:CaRel}) containing $N=7$ channels with additive Wiener noise. \textit{Grey bars} are fitted with the \textit{blue solid} curve and represent the distribution of IPIs obtained from the modeling of $622$ puffs. \textit{White bars} with \textit{black solid} fit are reproduced from the experiments reported in~\cite{Dickinson2012}.}   
\end{figure} 

\end{itemize}

There are other consequences for the fixed levels of \ip in the model.  Due to the fixed high level of [IP\textunderscript{3}] assumed, our model exhibits periods of oscillations lower than the wide range found in typical experimental studies. Although the high \ip assumption narrows the range of periods observed, it helps us to study the main characteristics of waves and oscillations phenomenologically. The variability of the periods can be achieved by incorporating more complex features of \ca signals such as stochasticity \cite{Skupin2008,Thurley2011} and array enhanced coherence resonance~\cite{Falcke2003} under dynamical [IP\textunderscript{3}] loads.

There are other characteristics of the emergent \ca waves and oscillations in our model that matches those reported in the extensive experimental literature~\cite{Bootman1997a,Marchant1999,Marchant2001,Callamaras1998}. 
In particular, Marchant and Parker~\cite{Marchant2001} have shown that an increase in [IP\textunderscript{3}] shortens intervals between Ca\textupperscript{2+} release events. Our model agrees with these experiments by exhibiting oscillations with frequency increasing with increase in [IP\textunderscript{3}]. The periods of oscillations are also consistent with R{\"{u}}ckl. et al.~\cite{Ruckl2015}. The amplitudes of oscillations are in a range of observed puffs and waves~\cite{Par1996}. Also, the effect of wave collision found experimentally is present in our model.

In this paper, a reduced description of clustered channels with a parametric representation of the dynamics of subunits of channels is present alongside membrane-level diffusion-enabled interactions.  While the simplifications introduced in the reduction scheme can lead to model behavior inconsistent with empirical observations, the dynamical systems analysis presented here as a means of linking phenomena at different scales also enables us to propose explanatory mechanisms that could be accommodated in more elaborate models.

See Supplemental Material for the extended bifurcation analysis of the single-cluster model (Subsec.~\ref{SC}), the figure illustrating alternans (Subsec.~\ref{CWO}), the change of \ip (Subsec.~\ref{ip3}) derived from diffusion equation. 

\begin{acknowledgments}
 S. B. would like to acknowledge Alex Vasilev for inspiration for this research. Also, Andrii Iakovliev for useful ideas and proofreading the manuscript. Current research has been funded by Erasmus Mundus ACTIVE grant and Faculty of Engineering and the Environment at the University of Southampton.     
\end{acknowledgments}

\appendix
\section{DYK reduction to the three-state model} \label{A1}
The kinetics governing the upper plane of DYK cube [Fig.~\ref{fig:DYKRud}~(b)] is formulated as the next set of differential equations

\begin{eqnarray}
\dot{x}_{110} & = & -x_{110}(b_{1}+a_{2}c+b_{5})+x_{010}a_{1}p+ \nonumber \\
&& +x_{111}b_{2}+x_{100}a_{5}c,\label{eq:110}\\
\dot{x}_{100} & = & -x_{100}(b_{1}+a_{2}c+a_{5}c)+x_{000}a_{1}p+ \nonumber \\
&&+x_{101}b_{2}+x_{110}b_{5},\label{eq:100}\\
\dot{x}_{101} & = & -x_{101}(b_{2}+a_{2}c+a_{5}c)+x_{100}a_{2}c+ \nonumber \\
&&+x_{001}a_{3}p+x_{111}b_{5},\label{eq:101}\\
\dot{x}_{111} & = & -x_{111}(b_{2}+b_{3}+b_{5})+x_{110}a_{2}c+ \nonumber \\
&&+x_{011}a_{3}p+x_{101}a_{5}c,\label{eq:111}
\end{eqnarray}
where $x_{110}+x_{100}+x_{101}+x_{111} = 1-y$, $y = x_{000}+x_{001}+x_{011}+x_{010}$. The fractions of subunits in state $ijk$ are denoted as $x_{ijk}$.

In the case of high [IP\textunderscript{3}] the IP\textunderscript{3} binding sites are saturated. Therefore, we can apply the condition of detailed balance between upper and lower planes of DYK: $a_ip x_{0jk} = b_ix_{1jk}$ for four sets of $i$, $j$, and $k$ such as $i=1$, $j=0,1$, $k=0$ and $j=3$, $j=0,1$, $k=1$. From these conditions we derive

\begin{eqnarray}
{x_{000}} & = & \frac{x_{100}b_1}{a_1p},\label{eq:det000}\\
{x_{010}} & = & \frac{x_{110}b_1}{a_1p},\label{eq:det010}\\
{x_{001}} & = & \frac{x_{101}b_3}{a_3p},\label{eq:det001}\\
{x_{011}} & = & \frac{x_{111}b_3}{a_3p}.\label{eq:det011}
\end{eqnarray} 

We substitute eqs.~(\ref{eq:det000})--(\ref{eq:det011}) into the system of eqs.~(\ref{eq:110})--(\ref{eq:111}). The resulting four state model appears as 
\begin{eqnarray}
\dot{x}_{110} & = & -x_{110}(a_{2}c+b_{5})+x_{111}b_{2}+x_{100}a_{5}c,\label{eq:110R}\\
\dot{x}_{100} & = & -x_{100}(a_{2}c+a_{5}c)+x_{101}b_{2}+x_{110}b_{5},\label{eq:100R}\\
\dot{x}_{101} & = & -x_{101}(b_{2}+a_{5}c)+x_{100}a_{2}c+x_{111}b_{5},\label{eq:101R}\\
\dot{x}_{111} & = & -x_{111}(b_{2}+b_{5})+x_{110}a_{2}c+x_{101}a_{5}c,\label{eq:111R}
\end{eqnarray}
where $x_{110}+x_{100}+x_{101}+x_{111}=1-y$ due to conservation of probability, $y=x_{010}+x_{000}+x_{001}+x_{011}$. 

Now we introduce a reduction of DYK similarly to Li and Rinzel~\cite{Li1994}. However, here we separate the upper and lower planes of the DYK cube [Fig.~\ref{fig:DYKRud}~(a)]. Considering the lower plane of the DYK cube in a steady state we obtain

\begin{eqnarray}
{x_{000}}&=&\frac{d_5d_4y}{(c+d_5)(c+d_4)}, \\ \label{eq:000_ss}
{x_{001}}&=&\frac{d_5cy}{(c+d_5)(c+d_4)}, \\ \label{eq:001_ss}
{x_{010}}&=&\frac{d_4cy}{(c+d_5)(c+d_4)}, \\ \label{eq:010_ss}
{x_{011}}&=&\frac{c^2y}{(c+d_5)(c+d_4)}, \label{eq:011_ss}
\end{eqnarray} 
where $d_i=b_i/a_i$, $i=\overline{1,5}$.

Similarly, the upper plane is in a steady state gives

\begin{eqnarray}
{x_{100}}&=&\frac{d_5d_2(1-y)}{(c+d_5)(c+d_2)}, \\ \label{eq:100_ss}
{x_{101}}&=&\frac{d_5c(1-y)}{(c+d_5)(c+d_2)}, \\ \label{eq:101_ss}
{x_{110}}&=&\frac{d_2c(1-y)}{(c+d_5)(c+d_2)}, \\ \label{eq:110_ss}
{x_{111}}&=&\frac{c^2(1-y)}{(c+d_5)(c+d_2)}. \label{eq:111_ss}
\end{eqnarray} 

To calculate the transition rates between these two groups we use steady state equations for upper plane (\ref{eq:000_ss})--(\ref{eq:011_ss}) and lower plane (\ref{eq:100_ss})--(\ref{eq:111_ss})

\begin{eqnarray}
K^-&=&a_1p(x_{000}+x_{010})+a_3p(x_{001}+x_{011})= \nonumber \\
&=&\frac{(a_3c+a_1d_4)py}{c+d_4}, \\
K^+&=&b_1(x_{100}+x_{110})+b_3(x_{101}+x_{111})= \nonumber \\
&=&\frac{(b_3c+b_1d_2)(1-y)}{c+d_2},
\end{eqnarray}

here we define transition rates between upper and lower planes of the DYK cube as
\begin{eqnarray}
\label{eq:c_pm}
C^+=\frac{b_3c+b_1d_2}{c+d_2},\ 
C^-=\frac{(a_3c+a_1d_4)p}{c+d_4},
\end{eqnarray}
which is used in (\ref{eq:y_2ip3}), the corresponding transition rates are given in Tab.~\ref{tab:IP3}.

\begin{table}[b]
	\caption{\label{tab:IP3}The parameters of the \ip dependent model adapted from~\cite{Ruckl2015,Ruckl2016}.}
	\begin{ruledtabular}
		\begin{tabular}{c c}
			Parameter & Value \\
			\hline
			First order rates, $\mu \text{M}^{-1}\times \text{s}^{-1}$  &   \\	
			\hline
			$a_1$ & $2$ \\
			$a_3$ & $4$ \\
			$a_4$ & $1$ \\
			\hline
			Second order rates, $\text{s}^{-1}$ & \\
			\hline
			$b_1$ & $2\times 10^{-3}$ \\
			$b_3$ & $8$ \\
			$b_4$ & $3.9\times 10^{-2}$ \\
			\hline
			Dissociation constants $d_i=b_i/a_i$, $\mu \text{M}$ & \\
			\hline
			$d_1$ & $0.001$ \\
			$d_3$ & $2$ \\
			$d_4$ & $0.039$ \\
		\end{tabular}
	\end{ruledtabular}
\end{table}

The results presented by the system of eqs.~(\ref{eq:110R})--(\ref{eq:111R}) are applicable to a subunit of a channel. In order to describe the behavior of a cluster of channels, similarly to ~\cite{Rudiger2014}, we introduce variables $a$, $g$, $h'$ and $z$ that represent fractions of channels in states $110$, $111$, $101$, and $100$, respectively. Thus including the \ip dependence the system (\ref{eq:100R})--(\ref{eq:111R}) transforms to

\begin{eqnarray}
\frac{\dif a}{\dif t} & = & k_{a}^{+}cz-k_{a}^{-}a+k_{i}^{-}g-\tilde{k}_{i}^{+}a,\label{eq:a_IP3}\\
\frac{\dif g}{\dif t} & = & k_{a}^{+}ch'-k_{a}^{-}g-k_{i}^{-}g+\tilde{k}_{i}^{+}a,\label{eq:g_IP3}\\
\frac{\dif h'}{\dif t} & = & k_a^-g - k_a^+ch' + k_i^+cz - k_i^-h',\label{eq:h'_IP3} \\
\frac{\dif y}{\dif t} & = & C^+(1-y)-C^-_\text{ch}y,\label{eq:y_IP3}
\end{eqnarray}
where $z=1-a-g-h'-y$ as sum of the fractions of all states to unity, $y$ defines the fraction of \ipr unbound channels (occupying the lower plane of the DYK cube), $C^+$ and $C^-_\text{ch}$ are given by eqs.~(\ref{eq:c_pm}) and (\ref{eq:c_m_ch}), respectively.

The kinetics of these variables is governed by rates $k_{a}^{\pm}$, $k_{i}^{\pm}$ and concentrations $c$ and $c_s$, where we substitute two parameters $k^+_i$ and $c_s$ as $\tilde{k}^+_i=k^+_ic_s$. These transition rates are obtained from the original DYK rates. Earlier simulations~\cite{Rudiger2014} showed that most of the reactions ($k_{a}^{-}$, $k_{i}^{+}$, and $k_{i}^{-}$) in the four-state model are obtained from $b_5$, $a_2$, and $b_2$, respectively. In contrast, $k_{a}^{+}$ rate has been constructed from the condition of channel opening (at least three subunits should be active). The probability of the single subunit activated is $P_\text{act}=a_{5}c/(a_{5}c+b_{5})$. According to the binomial distributions, the probability of none of the subunits is active is $P_{0}=(1-P_\text{act})^{4}$. If only one of four subunits is active and all remaining ones are inactive $P_{1}=4P_\text{act}(1-P_\text{act})^{3}$. There are $C_4^2=\frac{4!}{2!2!}=6$ ways for two of four subunits to be activated, thus, $P_{2}=6P_\text{act}^{2}(1-P_\text{act})^{2}$. Therefore, the transition to this state is possible only after the activation of two subunits. The probability that two subunits are active under the condition that the channel is closed is $P(2\arrowvert\{0,1,2\})=P_{2}/(P_{0}+P_{1}+P_{2})$.

We calculate $k_a^+$ as a transition rate from a closed channel with two active subunits to an open channel with three activated subunits. One of two remaining inactive subunits might be activated, thus, we obtain 
\begin{equation}
k_{a}^{+}=2a_{5}P(2\arrowvert\{0,1,2\}).\label{eq:ka+}
\end{equation}

Similarly to $k_a^+$ we derive $C^-_\text{ch}$ which requires four subunits to bind \ip~\cite{Taylor2016}

\begin{equation}
C^-_\text{ch} = C^- P_\text{IP}(3\arrowvert\{0,1,2,3\}), \label{eq:c_m_ch}
\end{equation}  
where $C^-$ is given by eq.~(\ref{eq:c_pm}), $P_\text{IP}(3\arrowvert\{0,1,2,3\}) = P^\text{IP}_3/(P^\text{IP}_0 + P^\text{IP}_1 + P^\text{IP}_2 + P^\text{IP}_3)$, similarly $P^\text{IP}_\text{act} = C^-/(C^+ + C^-)$, $P^\text{IP}_0 = (1-P^\text{IP}_\text{act})^{4}$, $P^\text{IP}_1 = 4 P^\text{IP}_\text{act}(1-P^\text{IP}_\text{act})^{3}$, $P^\text{IP}_2 = 6 (P^\text{IP}_\text{act})^2(1-P^\text{IP}_\text{act})^{2}$, $P^\text{IP}_3 = 4 (P^\text{IP}_\text{act})^3(1-P^\text{IP}_\text{act})$.

In order to reduce the number of variables, similarly to~\cite{Rudiger2014} we introduce a compound state $h$ which contains $g$ and $h'$ [Fig.~\ref{fig:DYKRud}~(d)].
The effective rates of transitions to this state are obtained from a detailed balance between $g$ and $h'$. We introduce the fraction of channels in the compound state as the sum of components $h=g+h'$, where from detailed balance 

\begin{eqnarray}
h&=&\frac{k_a^-}{k_a^+c}g+g=\frac{k_a^+c+k_a^-}{k_a^+c}g, \\
h&=&h'+\frac{k_a^-}{k_a^+c}h'=\frac{k_a^+c+k_a^-}{k_a^-}h'.
\end{eqnarray}
Considering that the rates of transition to compound state are 

\begin{eqnarray}
k_1&=&k_i^-\frac{k_a^+c}{k_a^+c+k_a^-}=k_i^-g_0, \label{eq:k1} \\
k_2&=&k_i^-\frac{k_a^-}{k_a^+c+k_a^-}=k_i^-(1-g_0), \label{eq:k2}
\end{eqnarray}
where $g_{0}=k_{a}^{+}c/(k_{a}^{+}c+k_{a}^{-})$. 

The resulting model reads as
\begin{eqnarray}
\frac{\dif a}{\dif t} & = & k_{a}^{+}(c)(1-a-h-y)-k_{a}^{-}a+k_{1}(c)h-\tilde{k}_{i}^{+}a \nonumber \\
&&\triangleq f_p(a,h,y,c),\label{eq:a_z}\\
\frac{\dif h}{\dif t} & = & k_{i}^{+}c(1-a-h-y)-k_{1}(c)h-k_{2}(c)h+\tilde{k}_{i}^{+}a \nonumber \\
&&\triangleq g_p(a,h,y,c),\label{eq:h_z}
\end{eqnarray}
which where $f_p$ and $g_p$ are used in eqs.~(\ref{eq:a_2ip3}), (\ref{eq:h_2ip3}).

The solution of eq.~(\ref{eq:y_IP3}) separately is 

\begin{equation}
y(t)=Ae^{-(C^++C^-)t}+\frac{C^+}{C^++C^-}.
\end{equation}

It follows that the equilibrium solution for the system is 

\begin{widetext}
	\begin{equation}
	y_{eq}=\frac{C^+}{C^++C^-}=\frac{(b_3c+b_1d_2)(c+d_4)}{(a_1d_4+a_3c)(c+d_2)p+(b_1d_2+b_3c)(c+d_4)}\overset{p\gg 1}{\longrightarrow}\frac{const}{p},\label{eq:IP3red}
	\end{equation}
\end{widetext}

in the case of  IP\textunderscript{3} saturation, we consider $p=$ [IP\textunderscript{3}] to be high. Taking that into account and assuming that waiting time is long enough $t\gg 1/(C^++C^-)$ we obtain $y\rightarrow 0$. Considering all the assumptions made before, we further operate with the three-variable system taking into account the summing fractions to unity $a+h+z=1$.

Using the assumption of high [IP\textunderscript{3}] from (\ref{eq:a_z}), (\ref{eq:h_z}), (\ref{eq:y_IP3}) we obtain the model illustrated in Fig.~\ref{fig:DYKRud}~(d) and given by eqs.~(\ref{eq:active}), (\ref{eq:closed}).

\section{Stochastic model} \label{A2}

In this appendix, we introduce the Langevin model of \ca release from a single cluster of channels. By introducing the noise terms into the model from eqs.~(\ref{eq:active}), (\ref{eq:closed}) we obtain 

\begin{eqnarray}
{\dif a} & = &  f(a,h,c)\dif t + A s_a\dif W,\label{eq:stoch_a}\\
{\dif h} & = &  g(a,h,c)\dif t + A s_h\dif W,\label{eq:stoch_h}
\end{eqnarray}  
where $\dif W$ are Wiener increments (generated by the Wiener process also known as Brownian motion~\cite{Oksendal2003}), $s_a$ and $s_h$ are the functions of the system state in the general case of multiplicative noise, but here for simplicity, we consider $s_a=s_h=1$ which corresponds to the additive noise case, $A=0.37$ is the amplitude of the noise. 

We integrate the system (\ref{eq:stoch_a}), (\ref{eq:stoch_h}) under $c=c_d$ assumption using the Euler-Maruyama method~\cite{Horchler2013a} with parameters given in the last column of Tab.~\ref{tab:RudOrig} and $N=7$. The results presented in Fig.~\ref{fig:stochastic} are obtained by analyzing the intervals between $622$ puffs using [Ca\textupperscript{2+}] traces of several thousands of seconds.

%


\begin{thebibliography}{54}%
	\makeatletter
	\providecommand \@ifxundefined [1]{%
		\@ifx{#1\undefined}
	}%
	\providecommand \@ifnum [1]{%
		\ifnum #1\expandafter \@firstoftwo
		\else \expandafter \@secondoftwo
		\fi
	}%
	\providecommand \@ifx [1]{%
		\ifx #1\expandafter \@firstoftwo
		\else \expandafter \@secondoftwo
		\fi
	}%
	\providecommand \natexlab [1]{#1}%
	\providecommand \enquote  [1]{``#1''}%
	\providecommand \bibnamefont  [1]{#1}%
	\providecommand \bibfnamefont [1]{#1}%
	\providecommand \citenamefont [1]{#1}%
	\providecommand \href@noop [0]{\@secondoftwo}%
	\providecommand \href [0]{\begingroup \@sanitize@url \@href}%
	\providecommand \@href[1]{\@@startlink{#1}\@@href}%
	\providecommand \@@href[1]{\endgroup#1\@@endlink}%
	\providecommand \@sanitize@url [0]{\catcode `\\12\catcode `\$12\catcode
		`\&12\catcode `\#12\catcode `\^12\catcode `\_12\catcode `\%12\relax}%
	\providecommand \@@startlink[1]{}%
	\providecommand \@@endlink[0]{}%
	\providecommand \url  [0]{\begingroup\@sanitize@url \@url }%
	\providecommand \@url [1]{\endgroup\@href {#1}{\urlprefix }}%
	\providecommand \urlprefix  [0]{URL }%
	\providecommand \Eprint [0]{\href }%
	\providecommand \doibase [0]{http://dx.doi.org/}%
	\providecommand \selectlanguage [0]{\@gobble}%
	\providecommand \bibinfo  [0]{\@secondoftwo}%
	\providecommand \bibfield  [0]{\@secondoftwo}%
	\providecommand \translation [1]{[#1]}%
	\providecommand \BibitemOpen [0]{}%
	\providecommand \bibitemStop [0]{}%
	\providecommand \bibitemNoStop [0]{.\EOS\space}%
	\providecommand \EOS [0]{\spacefactor3000\relax}%
	\providecommand \BibitemShut  [1]{\csname bibitem#1\endcsname}%
	\let\auto@bib@innerbib\@empty
	\bibitem [{\citenamefont {Berridge}(2012)}]{Berridge1997a}%
	\BibitemOpen
	\bibfield  {author} {\bibinfo {author} {\bibfnamefont {M.~J.}\ \bibnamefont
			{Berridge}},\ }\href {\doibase 10.1042/BST20110766} {\bibfield  {journal}
		{\bibinfo  {journal} {Biochem Soc T}\ }\textbf {\bibinfo {volume} {40}},\
		\bibinfo {pages} {297} (\bibinfo {year} {2012})}\BibitemShut {NoStop}%
	\bibitem [{\citenamefont {Mikoshiba}(2007)}]{Mikoshiba2007}%
	\BibitemOpen
	\bibfield  {author} {\bibinfo {author} {\bibfnamefont {K.}~\bibnamefont
			{Mikoshiba}},\ }\href {http://symposia.biochemistry.org/content/74/9}
	{\bibfield  {journal} {\bibinfo  {journal} {Biochem Soc Symp}\ }\textbf
		{\bibinfo {volume} {74}},\ \bibinfo {pages} {9} (\bibinfo {year}
		{2007})}\BibitemShut {NoStop}%
	\bibitem [{\citenamefont {Taufiq-Ur-Rahman}\ \emph {et~al.}(2009)\citenamefont
		{Taufiq-Ur-Rahman}, \citenamefont {Skupin}, \citenamefont {Falcke},\ and\
		\citenamefont {Taylor}}]{Taufiq-Ur-Rahman2009}%
	\BibitemOpen
	\bibfield  {author} {\bibinfo {author} {\bibnamefont {Taufiq-Ur-Rahman}},
		\bibinfo {author} {\bibfnamefont {A.}~\bibnamefont {Skupin}}, \bibinfo
		{author} {\bibfnamefont {M.}~\bibnamefont {Falcke}}, \ and\ \bibinfo {author}
		{\bibfnamefont {C.~W.}\ \bibnamefont {Taylor}},\ }\href {\doibase
		10.1038/nature07763} {\bibfield  {journal} {\bibinfo  {journal} {Nature}\
		}\textbf {\bibinfo {volume} {458}},\ \bibinfo {pages} {655} (\bibinfo {year}
		{2009})}\BibitemShut {NoStop}%
	\bibitem [{\citenamefont {Smith}\ \emph {et~al.}(2014)\citenamefont {Smith},
		\citenamefont {Swaminathan}, \citenamefont {Dickinson},\ and\ \citenamefont
		{Parker}}]{Smith2014}%
	\BibitemOpen
	\bibfield  {author} {\bibinfo {author} {\bibfnamefont {I.~F.}\ \bibnamefont
			{Smith}}, \bibinfo {author} {\bibfnamefont {D.}~\bibnamefont {Swaminathan}},
		\bibinfo {author} {\bibfnamefont {G.~D.}\ \bibnamefont {Dickinson}}, \ and\
		\bibinfo {author} {\bibfnamefont {I.}~\bibnamefont {Parker}},\ }\href
	{\doibase 10.1016/j.bpj.2014.05.051} {\bibfield  {journal} {\bibinfo
			{journal} {Biophys J}\ }\textbf {\bibinfo {volume} {107}},\ \bibinfo {pages}
		{834} (\bibinfo {year} {2014})}\BibitemShut {NoStop}%
	\bibitem [{\citenamefont {Dickinson}\ \emph {et~al.}(2012)\citenamefont
		{Dickinson}, \citenamefont {Swaminathan},\ and\ \citenamefont
		{Parker}}]{Dickinson2012}%
	\BibitemOpen
	\bibfield  {author} {\bibinfo {author} {\bibfnamefont {G.~D.}\ \bibnamefont
			{Dickinson}}, \bibinfo {author} {\bibfnamefont {D.}~\bibnamefont
			{Swaminathan}}, \ and\ \bibinfo {author} {\bibfnamefont {I.}~\bibnamefont
			{Parker}},\ }\href {\doibase 10.1016/j.bpj.2012.03.029} {\bibfield  {journal}
		{\bibinfo  {journal} {Biophys J}\ }\textbf {\bibinfo {volume} {102}},\
		\bibinfo {pages} {1826} (\bibinfo {year} {2012})}\BibitemShut {NoStop}%
	\bibitem [{\citenamefont {Berridge}(2007)}]{Berridge2007}%
	\BibitemOpen
	\bibfield  {author} {\bibinfo {author} {\bibfnamefont {M.~J.}\ \bibnamefont
			{Berridge}},\ }\href {\doibase 10.1042/BSS0740001} {\bibfield  {journal}
		{\bibinfo  {journal} {Biochem Soc Symp}\ }\textbf {\bibinfo {volume} {74}},\
		\bibinfo {pages} {1} (\bibinfo {year} {2007})}\BibitemShut {NoStop}%
	\bibitem [{\citenamefont {{De Young}}\ and\ \citenamefont
		{Keizer}(1992)}]{DeYoung1992}%
	\BibitemOpen
	\bibfield  {author} {\bibinfo {author} {\bibfnamefont {G.~W.}\ \bibnamefont
			{{De Young}}}\ and\ \bibinfo {author} {\bibfnamefont {J.}~\bibnamefont
			{Keizer}},\ }\href {\doibase 10.1073/pnas.89.20.9895} {\bibfield  {journal}
		{\bibinfo  {journal} {P Natl A Sci USA}\ }\textbf {\bibinfo {volume} {89}},\
		\bibinfo {pages} {9895} (\bibinfo {year} {1992})}\BibitemShut {NoStop}%
	\bibitem [{\citenamefont {Li}\ and\ \citenamefont {Rinzel}(1994)}]{Li1994}%
	\BibitemOpen
	\bibfield  {author} {\bibinfo {author} {\bibfnamefont {Y.-X.}\ \bibnamefont
			{Li}}\ and\ \bibinfo {author} {\bibfnamefont {J.}~\bibnamefont {Rinzel}},\
	}\href {\doibase 10.1006/jtbi.1994.1041} {\bibfield  {journal} {\bibinfo
			{journal} {J Theor Biol}\ }\textbf {\bibinfo {volume} {166}},\ \bibinfo
		{pages} {461} (\bibinfo {year} {1994})}\BibitemShut {NoStop}%
	\bibitem [{\citenamefont {Politi}\ \emph {et~al.}(2006)\citenamefont {Politi},
		\citenamefont {Gaspers}, \citenamefont {Thomas},\ and\ \citenamefont
		{H{\"{o}}fer}}]{Politi2006}%
	\BibitemOpen
	\bibfield  {author} {\bibinfo {author} {\bibfnamefont {A.}~\bibnamefont
			{Politi}}, \bibinfo {author} {\bibfnamefont {L.~D.}\ \bibnamefont {Gaspers}},
		\bibinfo {author} {\bibfnamefont {A.~P.}\ \bibnamefont {Thomas}}, \ and\
		\bibinfo {author} {\bibfnamefont {T.}~\bibnamefont {H{\"{o}}fer}},\ }\href
	{\doibase 10.1529/biophysj.105.072249} {\bibfield  {journal} {\bibinfo
			{journal} {Biophys J}\ }\textbf {\bibinfo {volume} {90}},\ \bibinfo {pages}
		{3120} (\bibinfo {year} {2006})}\BibitemShut {NoStop}%
	\bibitem [{\citenamefont {Smedler}\ and\ \citenamefont
		{Uhl{\'{e}}n}(2014)}]{Smedler2014}%
	\BibitemOpen
	\bibfield  {author} {\bibinfo {author} {\bibfnamefont {E.}~\bibnamefont
			{Smedler}}\ and\ \bibinfo {author} {\bibfnamefont {P.}~\bibnamefont
			{Uhl{\'{e}}n}},\ }\href
	{https://www.sciencedirect.com/science/article/pii/S0304416513005163?via{\%}3Dihub}
	{\bibfield  {journal} {\bibinfo  {journal} {BBA-Gen Subjects}\ }\textbf
		{\bibinfo {volume} {1840}},\ \bibinfo {pages} {964} (\bibinfo {year}
		{2014})}\BibitemShut {NoStop}%
	\bibitem [{\citenamefont {Falcke}(2003)}]{Falcke2003}%
	\BibitemOpen
	\bibfield  {author} {\bibinfo {author} {\bibfnamefont {M.}~\bibnamefont
			{Falcke}},\ }\href {\doibase 10.1016/S0006-3495(03)74831-0} {\bibfield
		{journal} {\bibinfo  {journal} {Biophys J}\ }\textbf {\bibinfo {volume}
			{84}},\ \bibinfo {pages} {42} (\bibinfo {year} {2003})}\BibitemShut {NoStop}%
	\bibitem [{\citenamefont {Skupin}\ \emph {et~al.}(2008)\citenamefont {Skupin},
		\citenamefont {Kettenmann}, \citenamefont {Winkler}, \citenamefont
		{Wartenberg}, \citenamefont {Sauer}, \citenamefont {Tovey}, \citenamefont
		{Taylor},\ and\ \citenamefont {Falcke}}]{Skupin2008}%
	\BibitemOpen
	\bibfield  {author} {\bibinfo {author} {\bibfnamefont {A.}~\bibnamefont
			{Skupin}}, \bibinfo {author} {\bibfnamefont {H.}~\bibnamefont {Kettenmann}},
		\bibinfo {author} {\bibfnamefont {U.}~\bibnamefont {Winkler}}, \bibinfo
		{author} {\bibfnamefont {M.}~\bibnamefont {Wartenberg}}, \bibinfo {author}
		{\bibfnamefont {H.}~\bibnamefont {Sauer}}, \bibinfo {author} {\bibfnamefont
			{S.~C.}\ \bibnamefont {Tovey}}, \bibinfo {author} {\bibfnamefont {C.~W.}\
			\bibnamefont {Taylor}}, \ and\ \bibinfo {author} {\bibfnamefont
			{M.}~\bibnamefont {Falcke}},\ }\href {\doibase 10.1529/biophysj.107.119495}
	{\bibfield  {journal} {\bibinfo  {journal} {Biophys J}\ }\textbf {\bibinfo
			{volume} {94}},\ \bibinfo {pages} {2404} (\bibinfo {year}
		{2008})}\BibitemShut {NoStop}%
	\bibitem [{\citenamefont {Thurley}\ \emph {et~al.}(2011)\citenamefont
		{Thurley}, \citenamefont {Smith}, \citenamefont {Tovey}, \citenamefont
		{Taylor}, \citenamefont {Parker},\ and\ \citenamefont
		{Falcke}}]{Thurley2011}%
	\BibitemOpen
	\bibfield  {author} {\bibinfo {author} {\bibfnamefont {K.}~\bibnamefont
			{Thurley}}, \bibinfo {author} {\bibfnamefont {I.~F.}\ \bibnamefont {Smith}},
		\bibinfo {author} {\bibfnamefont {S.~C.}\ \bibnamefont {Tovey}}, \bibinfo
		{author} {\bibfnamefont {C.~W.}\ \bibnamefont {Taylor}}, \bibinfo {author}
		{\bibfnamefont {I.}~\bibnamefont {Parker}}, \ and\ \bibinfo {author}
		{\bibfnamefont {M.}~\bibnamefont {Falcke}},\ }\href {\doibase
		10.1016/j.bpj.2011.10.030} {\bibfield  {journal} {\bibinfo  {journal}
			{Biophys J}\ }\textbf {\bibinfo {volume} {101}},\ \bibinfo {pages} {2638}
		(\bibinfo {year} {2011})}\BibitemShut {NoStop}%
	\bibitem [{\citenamefont {Kupferman}\ \emph {et~al.}(1997)\citenamefont
		{Kupferman}, \citenamefont {Mitra}, \citenamefont {Hohenberg},\ and\
		\citenamefont {Wang}}]{Kupferman1997}%
	\BibitemOpen
	\bibfield  {author} {\bibinfo {author} {\bibfnamefont {R.}~\bibnamefont
			{Kupferman}}, \bibinfo {author} {\bibfnamefont {P.~P.}\ \bibnamefont
			{Mitra}}, \bibinfo {author} {\bibfnamefont {P.~C.}\ \bibnamefont
			{Hohenberg}}, \ and\ \bibinfo {author} {\bibfnamefont {S.~S.-H.}\
			\bibnamefont {Wang}},\ }\href {\doibase 10.1016/S0006-3495(97)78888-X}
	{\bibfield  {journal} {\bibinfo  {journal} {Biophys J}\ }\textbf {\bibinfo
			{volume} {72}},\ \bibinfo {pages} {2430} (\bibinfo {year}
		{1997})}\BibitemShut {NoStop}%
	\bibitem [{\citenamefont {Atri}\ \emph {et~al.}(1993)\citenamefont {Atri},
		\citenamefont {Amundson}, \citenamefont {Clapham},\ and\ \citenamefont
		{Sneyd}}]{Atri1993}%
	\BibitemOpen
	\bibfield  {author} {\bibinfo {author} {\bibfnamefont {A.}~\bibnamefont
			{Atri}}, \bibinfo {author} {\bibfnamefont {J.}~\bibnamefont {Amundson}},
		\bibinfo {author} {\bibfnamefont {D.}~\bibnamefont {Clapham}}, \ and\
		\bibinfo {author} {\bibfnamefont {J.}~\bibnamefont {Sneyd}},\ }\href
	{\doibase 10.1016/S0006-3495(93)81191-3} {\bibfield  {journal} {\bibinfo
			{journal} {Biophys J}\ }\textbf {\bibinfo {volume} {65}},\ \bibinfo {pages}
		{1727} (\bibinfo {year} {1993})}\BibitemShut {NoStop}%
	\bibitem [{\citenamefont {Sneyd}\ \emph {et~al.}(1998)\citenamefont {Sneyd},
		\citenamefont {Dale},\ and\ \citenamefont {Duffy}}]{Sneyd1998}%
	\BibitemOpen
	\bibfield  {author} {\bibinfo {author} {\bibfnamefont {J.}~\bibnamefont
			{Sneyd}}, \bibinfo {author} {\bibfnamefont {P.~D.}\ \bibnamefont {Dale}}, \
		and\ \bibinfo {author} {\bibfnamefont {A.}~\bibnamefont {Duffy}},\ }\href
	{https://epubs.siam.org/doi/10.1137/S0036139996305074} {\bibfield  {journal}
		{\bibinfo  {journal} {SIAM J Appl Math}\ }\textbf {\bibinfo {volume} {58}},\
		\bibinfo {pages} {1178} (\bibinfo {year} {1998})}\BibitemShut {NoStop}%
	\bibitem [{\citenamefont {Dupont}\ and\ \citenamefont
		{Goldbeter}(1994)}]{Dupont1994}%
	\BibitemOpen
	\bibfield  {author} {\bibinfo {author} {\bibfnamefont {G.}~\bibnamefont
			{Dupont}}\ and\ \bibinfo {author} {\bibfnamefont {A.}~\bibnamefont
			{Goldbeter}},\ }\href {\doibase 10.1016/S0006-3495(94)80705-2} {\bibfield
		{journal} {\bibinfo  {journal} {Biophys J}\ }\textbf {\bibinfo {volume}
			{67}},\ \bibinfo {pages} {2191} (\bibinfo {year} {1994})}\BibitemShut
	{NoStop}%
	\bibitem [{\citenamefont {Calabrese}\ \emph {et~al.}(2010)\citenamefont
		{Calabrese}, \citenamefont {Fraiman}, \citenamefont {Zysman},\ and\
		\citenamefont {{Ponce Dawson}}}]{Calabrese2010}%
	\BibitemOpen
	\bibfield  {author} {\bibinfo {author} {\bibfnamefont {A.}~\bibnamefont
			{Calabrese}}, \bibinfo {author} {\bibfnamefont {D.}~\bibnamefont {Fraiman}},
		\bibinfo {author} {\bibfnamefont {D.}~\bibnamefont {Zysman}}, \ and\ \bibinfo
		{author} {\bibfnamefont {S.}~\bibnamefont {{Ponce Dawson}}},\ }\href
	{\doibase 10.1103/PhysRevE.82.031910} {\bibfield  {journal} {\bibinfo
			{journal} {Phys Rev E}\ }\textbf {\bibinfo {volume} {82}},\ \bibinfo {pages}
		{031910} (\bibinfo {year} {2010})}\BibitemShut {NoStop}%
	\bibitem [{\citenamefont {Keizer}\ and\ \citenamefont
		{Smith}(1998)}]{Keizer1998}%
	\BibitemOpen
	\bibfield  {author} {\bibinfo {author} {\bibfnamefont {J.}~\bibnamefont
			{Keizer}}\ and\ \bibinfo {author} {\bibfnamefont {G.~D.}\ \bibnamefont
			{Smith}},\ }\href {\doibase 10.1016/S0301-4622(98)00125-2} {\bibfield
		{journal} {\bibinfo  {journal} {Biophys Chem}\ }\textbf {\bibinfo {volume}
			{72}},\ \bibinfo {pages} {87} (\bibinfo {year} {1998})}\BibitemShut {NoStop}%
	\bibitem [{\citenamefont {Sneyd}\ and\ \citenamefont
		{Sherratt}(1997)}]{Sneyd1997}%
	\BibitemOpen
	\bibfield  {author} {\bibinfo {author} {\bibfnamefont {J.}~\bibnamefont
			{Sneyd}}\ and\ \bibinfo {author} {\bibfnamefont {J.}~\bibnamefont
			{Sherratt}},\ }\href {\doibase 10.1137/S0036139995286035} {\bibfield
		{journal} {\bibinfo  {journal} {SIAM J Appl Math}\ }\textbf {\bibinfo
			{volume} {57}},\ \bibinfo {pages} {73} (\bibinfo {year} {1997})}\BibitemShut
	{NoStop}%
	\bibitem [{\citenamefont {Dawson}\ \emph {et~al.}(1999)\citenamefont {Dawson},
		\citenamefont {Keizer},\ and\ \citenamefont {Pearson}}]{Dawson1999}%
	\BibitemOpen
	\bibfield  {author} {\bibinfo {author} {\bibfnamefont {S.~P.}\ \bibnamefont
			{Dawson}}, \bibinfo {author} {\bibfnamefont {J.}~\bibnamefont {Keizer}}, \
		and\ \bibinfo {author} {\bibfnamefont {J.~E.}\ \bibnamefont {Pearson}},\
	}\href {\doibase 10.1073/pnas.96.11.6060} {\bibfield  {journal} {\bibinfo
			{journal} {P Natl A Sci USA}\ }\textbf {\bibinfo {volume} {96}},\ \bibinfo
		{pages} {6060} (\bibinfo {year} {1999})}\BibitemShut {NoStop}%
	\bibitem [{\citenamefont {Coombes}\ and\ \citenamefont
		{Timofeeva}(2003)}]{Coombes2003}%
	\BibitemOpen
	\bibfield  {author} {\bibinfo {author} {\bibfnamefont {S.}~\bibnamefont
			{Coombes}}\ and\ \bibinfo {author} {\bibfnamefont {Y.}~\bibnamefont
			{Timofeeva}},\ }\href {\doibase 10.1103/PhysRevE.68.021915} {\bibfield
		{journal} {\bibinfo  {journal} {Phys Rev E}\ }\textbf {\bibinfo {volume}
			{68}},\ \bibinfo {pages} {021915} (\bibinfo {year} {2003})}\BibitemShut
	{NoStop}%
	\bibitem [{\citenamefont {R{\"{u}}diger}(2014{\natexlab{a}})}]{Rudiger2014}%
	\BibitemOpen
	\bibfield  {author} {\bibinfo {author} {\bibfnamefont {S.}~\bibnamefont
			{R{\"{u}}diger}},\ }\href {\doibase 10.1103/PhysRevE.89.062717} {\bibfield
		{journal} {\bibinfo  {journal} {Phys Rev E}\ }\textbf {\bibinfo {volume}
			{89}},\ \bibinfo {pages} {062717} (\bibinfo {year}
		{2014}{\natexlab{a}})}\BibitemShut {NoStop}%
	\bibitem [{\citenamefont {R{\"{u}}ckl}\ \emph {et~al.}(2015)\citenamefont
		{R{\"{u}}ckl}, \citenamefont {Parker}, \citenamefont {Marchant},
		\citenamefont {Nagaiah}, \citenamefont {Johenning},\ and\ \citenamefont
		{R{\"{u}}diger}}]{Ruckl2015}%
	\BibitemOpen
	\bibfield  {author} {\bibinfo {author} {\bibfnamefont {M.}~\bibnamefont
			{R{\"{u}}ckl}}, \bibinfo {author} {\bibfnamefont {I.}~\bibnamefont {Parker}},
		\bibinfo {author} {\bibfnamefont {J.~S.}\ \bibnamefont {Marchant}}, \bibinfo
		{author} {\bibfnamefont {C.}~\bibnamefont {Nagaiah}}, \bibinfo {author}
		{\bibfnamefont {F.~W.}\ \bibnamefont {Johenning}}, \ and\ \bibinfo {author}
		{\bibfnamefont {S.}~\bibnamefont {R{\"{u}}diger}},\ }\href {\doibase
		10.1371/journal.pcbi.1003965} {\bibfield  {journal} {\bibinfo  {journal}
			{PLOS Comput Biol}\ }\textbf {\bibinfo {volume} {11}},\ \bibinfo {pages}
		{e1003965} (\bibinfo {year} {2015})}\BibitemShut {NoStop}%
	\bibitem [{\citenamefont {R{\"{u}}ckl}\ and\ \citenamefont
		{R{\"{u}}diger}(2016)}]{Ruckl2016}%
	\BibitemOpen
	\bibfield  {author} {\bibinfo {author} {\bibfnamefont {M.}~\bibnamefont
			{R{\"{u}}ckl}}\ and\ \bibinfo {author} {\bibfnamefont {S.}~\bibnamefont
			{R{\"{u}}diger}},\ }\href {\doibase 10.1140/epje/i2016-16108-4} {\bibfield
		{journal} {\bibinfo  {journal} {Eur Phys J E}\ }\textbf {\bibinfo {volume}
			{39}},\ \bibinfo {pages} {1} (\bibinfo {year} {2016})}\BibitemShut {NoStop}%
	\bibitem [{\citenamefont {Taylor}\ and\ \citenamefont
		{Konieczny}(2016)}]{Taylor2016}%
	\BibitemOpen
	\bibfield  {author} {\bibinfo {author} {\bibfnamefont {C.~W.}\ \bibnamefont
			{Taylor}}\ and\ \bibinfo {author} {\bibfnamefont {V.}~\bibnamefont
			{Konieczny}},\ }\href
	{http://stke.sciencemag.org/content/sigtrans/9/422/pe1.full.pdf} {\bibfield
		{journal} {\bibinfo  {journal} {Sci Signal}\ }\textbf {\bibinfo {volume}
			{9}},\ \bibinfo {pages} {2} (\bibinfo {year} {2016})}\BibitemShut {NoStop}%
	\bibitem [{\citenamefont {Qu}\ \emph {et~al.}(2016)\citenamefont {Qu},
		\citenamefont {Liu},\ and\ \citenamefont {Nivala}}]{Qu2016}%
	\BibitemOpen
	\bibfield  {author} {\bibinfo {author} {\bibfnamefont {Z.}~\bibnamefont
			{Qu}}, \bibinfo {author} {\bibfnamefont {M.~B.}\ \bibnamefont {Liu}}, \ and\
		\bibinfo {author} {\bibfnamefont {M.}~\bibnamefont {Nivala}},\ }\href
	{\doibase 10.1038/srep35625} {\bibfield  {journal} {\bibinfo  {journal} {Sci
				Rep}\ }\textbf {\bibinfo {volume} {6}},\ \bibinfo {pages} {35625} (\bibinfo
		{year} {2016})}\BibitemShut {NoStop}%
	\bibitem [{\citenamefont {Groff}\ and\ \citenamefont
		{Smith}(2008)}]{Groff2008}%
	\BibitemOpen
	\bibfield  {author} {\bibinfo {author} {\bibfnamefont {J.~R.}\ \bibnamefont
			{Groff}}\ and\ \bibinfo {author} {\bibfnamefont {G.~D.}\ \bibnamefont
			{Smith}},\ }\href {\doibase 10.1529/biophysj.107.119982} {\bibfield
		{journal} {\bibinfo  {journal} {Biophys J}\ }\textbf {\bibinfo {volume}
			{95}},\ \bibinfo {pages} {135} (\bibinfo {year} {2008})}\BibitemShut
	{NoStop}%
	\bibitem [{\citenamefont {Hinch}(2004)}]{Hinch2004}%
	\BibitemOpen
	\bibfield  {author} {\bibinfo {author} {\bibfnamefont {R.}~\bibnamefont
			{Hinch}},\ }\href {\doibase 10.1016/S0006-3495(04)74203-4} {\bibfield
		{journal} {\bibinfo  {journal} {Biophys J}\ }\textbf {\bibinfo {volume}
			{86}},\ \bibinfo {pages} {1293} (\bibinfo {year} {2004})}\BibitemShut
	{NoStop}%
	\bibitem [{\citenamefont {Wiltgen}\ \emph {et~al.}(2014)\citenamefont
		{Wiltgen}, \citenamefont {Dickinson}, \citenamefont {Swaminathan},\ and\
		\citenamefont {Parker}}]{Wiltgen2014}%
	\BibitemOpen
	\bibfield  {author} {\bibinfo {author} {\bibfnamefont {S.~M.}\ \bibnamefont
			{Wiltgen}}, \bibinfo {author} {\bibfnamefont {G.~D.}\ \bibnamefont
			{Dickinson}}, \bibinfo {author} {\bibfnamefont {D.}~\bibnamefont
			{Swaminathan}}, \ and\ \bibinfo {author} {\bibfnamefont {I.}~\bibnamefont
			{Parker}},\ }\href {\doibase 10.1016/j.ceca.2014.06.005} {\bibfield
		{journal} {\bibinfo  {journal} {Cell Calcium}\ }\textbf {\bibinfo {volume}
			{56}},\ \bibinfo {pages} {157} (\bibinfo {year} {2014})}\BibitemShut
	{NoStop}%
	\bibitem [{\citenamefont {Laver}\ \emph {et~al.}(2013)\citenamefont {Laver},
		\citenamefont {Kong}, \citenamefont {Imtiaz},\ and\ \citenamefont
		{Cannell}}]{Laver2013}%
	\BibitemOpen
	\bibfield  {author} {\bibinfo {author} {\bibfnamefont {D.~R.}\ \bibnamefont
			{Laver}}, \bibinfo {author} {\bibfnamefont {C.~H.~T.}\ \bibnamefont {Kong}},
		\bibinfo {author} {\bibfnamefont {M.~S.}\ \bibnamefont {Imtiaz}}, \ and\
		\bibinfo {author} {\bibfnamefont {M.~B.}\ \bibnamefont {Cannell}},\ }\href
	{https://www.sciencedirect.com/science/article/pii/S002228281200377X?via{\%}3Dihub}
	{\bibfield  {journal} {\bibinfo  {journal} {J Mol Cell Cardiol}\ }\textbf
		{\bibinfo {volume} {54}},\ \bibinfo {pages} {98} (\bibinfo {year}
		{2013})}\BibitemShut {NoStop}%
	\bibitem [{\citenamefont {R{\"{u}}diger}\ \emph {et~al.}(2012)\citenamefont
		{R{\"{u}}diger}, \citenamefont {Jung},\ and\ \citenamefont
		{Shuai}}]{Rudiger2012}%
	\BibitemOpen
	\bibfield  {author} {\bibinfo {author} {\bibfnamefont {S.}~\bibnamefont
			{R{\"{u}}diger}}, \bibinfo {author} {\bibfnamefont {P.}~\bibnamefont {Jung}},
		\ and\ \bibinfo {author} {\bibfnamefont {J.-W.}\ \bibnamefont {Shuai}},\
	}\href {\doibase 10.1371/journal.pcbi.1002485} {\bibfield  {journal}
		{\bibinfo  {journal} {PLOS Comput Biol}\ }\textbf {\bibinfo {volume} {8}},\
		\bibinfo {pages} {e1002485} (\bibinfo {year} {2012})}\BibitemShut {NoStop}%
	\bibitem [{\citenamefont {Bezprozvanny}\ \emph {et~al.}(1991)\citenamefont
		{Bezprozvanny}, \citenamefont {Watras},\ and\ \citenamefont
		{Ehrlich}}]{Bezprozvanny1991}%
	\BibitemOpen
	\bibfield  {author} {\bibinfo {author} {\bibfnamefont {I.}~\bibnamefont
			{Bezprozvanny}}, \bibinfo {author} {\bibfnamefont {J.}~\bibnamefont
			{Watras}}, \ and\ \bibinfo {author} {\bibfnamefont {B.~E.}\ \bibnamefont
			{Ehrlich}},\ }\href {\doibase 10.1038/351751a0} {\bibfield  {journal}
		{\bibinfo  {journal} {Nature}\ }\textbf {\bibinfo {volume} {351}},\ \bibinfo
		{pages} {751} (\bibinfo {year} {1991})}\BibitemShut {NoStop}%
	\bibitem [{\citenamefont {R{\"{u}}diger}\ \emph {et~al.}(2010)\citenamefont
		{R{\"{u}}diger}, \citenamefont {Shuai},\ and\ \citenamefont
		{Sokolov}}]{Rudiger}%
	\BibitemOpen
	\bibfield  {author} {\bibinfo {author} {\bibfnamefont {S.}~\bibnamefont
			{R{\"{u}}diger}}, \bibinfo {author} {\bibfnamefont {J.~W.}\ \bibnamefont
			{Shuai}}, \ and\ \bibinfo {author} {\bibfnamefont {I.~M.}\ \bibnamefont
			{Sokolov}},\ }\href {\doibase 10.1103/PhysRevLett.105.048103} {\bibfield
		{journal} {\bibinfo  {journal} {Phys Rev Lett}\ }\textbf {\bibinfo {volume}
			{105}},\ \bibinfo {pages} {048103} (\bibinfo {year} {2010})}\BibitemShut
	{NoStop}%
	\bibitem [{\citenamefont {Shuai}\ \emph {et~al.}(2008)\citenamefont {Shuai},
		\citenamefont {Pearson},\ and\ \citenamefont {Parker}}]{Shuai2008}%
	\BibitemOpen
	\bibfield  {author} {\bibinfo {author} {\bibfnamefont {J.}~\bibnamefont
			{Shuai}}, \bibinfo {author} {\bibfnamefont {J.~E.}\ \bibnamefont {Pearson}},
		\ and\ \bibinfo {author} {\bibfnamefont {I.}~\bibnamefont {Parker}},\ }\href
	{\doibase 10.1529/biophysj.108.137182} {\bibfield  {journal} {\bibinfo
			{journal} {Biophys J}\ }\textbf {\bibinfo {volume} {95}},\ \bibinfo {pages}
		{3738} (\bibinfo {year} {2008})}\BibitemShut {NoStop}%
	\bibitem [{\citenamefont {Dhooge}\ \emph {et~al.}(2008)\citenamefont {Dhooge},
		\citenamefont {Govaerts}, \citenamefont {Kuznetsov}, \citenamefont {Meijer},\
		and\ \citenamefont {{Sautois. B.}}}]{Dhooge2008}%
	\BibitemOpen
	\bibfield  {author} {\bibinfo {author} {\bibfnamefont {A.}~\bibnamefont
			{Dhooge}}, \bibinfo {author} {\bibfnamefont {W.}~\bibnamefont {Govaerts}},
		\bibinfo {author} {\bibfnamefont {Y.~A.}\ \bibnamefont {Kuznetsov}}, \bibinfo
		{author} {\bibfnamefont {H.~G.}\ \bibnamefont {Meijer}}, \ and\ \bibinfo
		{author} {\bibnamefont {{Sautois. B.}}},\ }\href {\doibase
		10.1080/13873950701742754} {\bibfield  {journal} {\bibinfo  {journal} {Math
				Comp Model Dyn}\ }\textbf {\bibinfo {volume} {14}},\ \bibinfo {pages} {147}
		(\bibinfo {year} {2008})}\BibitemShut {NoStop}%
	\bibitem [{\citenamefont {Kuznetsov}(2000)}]{Kuznetsov1998}%
	\BibitemOpen
	\bibfield  {author} {\bibinfo {author} {\bibfnamefont {Y.~A.}\ \bibnamefont
			{Kuznetsov}},\ }\href {\doibase 10.1007/b98848} {\emph {\bibinfo {title}
			{{Elements of Applied Bifurcation Theory}}}},\ edited by\ \bibinfo {editor}
	{\bibfnamefont {J.~E.}\ \bibnamefont {Madsen}}\ and\ \bibinfo {editor}
	{\bibfnamefont {L.}~\bibnamefont {Sirovich}}\ (\bibinfo  {publisher}
	{Springer},\ \bibinfo {year} {2000})\ pp.\ \bibinfo {pages}
	{91--112}\BibitemShut {NoStop}%
	\bibitem [{\citenamefont {Callamaras}\ \emph {et~al.}(1998)\citenamefont
		{Callamaras}, \citenamefont {Marchant}, \citenamefont {Sun},\ and\
		\citenamefont {Parker}}]{Callamaras1998}%
	\BibitemOpen
	\bibfield  {author} {\bibinfo {author} {\bibfnamefont {N.}~\bibnamefont
			{Callamaras}}, \bibinfo {author} {\bibfnamefont {J.~S.}\ \bibnamefont
			{Marchant}}, \bibinfo {author} {\bibfnamefont {X.-P.}\ \bibnamefont {Sun}}, \
		and\ \bibinfo {author} {\bibfnamefont {I.}~\bibnamefont {Parker}},\ }\href
	{\doibase 10.1111/j.1469-7793.1998.081bo.x} {\bibfield  {journal} {\bibinfo
			{journal} {J Physiol}\ }\textbf {\bibinfo {volume} {509}},\ \bibinfo {pages}
		{81} (\bibinfo {year} {1998})}\BibitemShut {NoStop}%
	\bibitem [{\citenamefont {Allbritton}\ \emph {et~al.}(1992)\citenamefont
		{Allbritton}, \citenamefont {Meyer},\ and\ \citenamefont
		{Stryer}}]{Allbritton1992}%
	\BibitemOpen
	\bibfield  {author} {\bibinfo {author} {\bibfnamefont {N.}~\bibnamefont
			{Allbritton}}, \bibinfo {author} {\bibfnamefont {T.}~\bibnamefont {Meyer}}, \
		and\ \bibinfo {author} {\bibfnamefont {L.}~\bibnamefont {Stryer}},\ }\href
	{\doibase 10.1126/science.1465619} {\bibfield  {journal} {\bibinfo  {journal}
			{Science}\ }\textbf {\bibinfo {volume} {258}},\ \bibinfo {pages} {1812}
		(\bibinfo {year} {1992})}\BibitemShut {NoStop}%
	\bibitem [{\citenamefont {R{\"{u}}diger}(2014{\natexlab{b}})}]{Rudiger2014a}%
	\BibitemOpen
	\bibfield  {author} {\bibinfo {author} {\bibfnamefont {S.}~\bibnamefont
			{R{\"{u}}diger}},\ }\href {\doibase 10.1016/j.physrep.2013.09.002} {\bibfield
		{journal} {\bibinfo  {journal} {Phys Rep}\ }\textbf {\bibinfo {volume}
			{534}},\ \bibinfo {pages} {39} (\bibinfo {year}
		{2014}{\natexlab{b}})}\BibitemShut {NoStop}%
	\bibitem [{\citenamefont {Dickinson}\ \emph {et~al.}(2016)\citenamefont
		{Dickinson}, \citenamefont {Ellefsen}, \citenamefont {Dawson}, \citenamefont
		{Pearson},\ and\ \citenamefont {Parker}}]{Dickinson2016}%
	\BibitemOpen
	\bibfield  {author} {\bibinfo {author} {\bibfnamefont {G.~D.}\ \bibnamefont
			{Dickinson}}, \bibinfo {author} {\bibfnamefont {K.~L.}\ \bibnamefont
			{Ellefsen}}, \bibinfo {author} {\bibfnamefont {S.~P.}\ \bibnamefont
			{Dawson}}, \bibinfo {author} {\bibfnamefont {J.~E.}\ \bibnamefont {Pearson}},
		\ and\ \bibinfo {author} {\bibfnamefont {I.}~\bibnamefont {Parker}},\ }\href
	{\doibase 10.1126/scisignal.aag1625} {\bibfield  {journal} {\bibinfo
			{journal} {Sci Signal}\ }\textbf {\bibinfo {volume} {9}},\ \bibinfo {pages}
		{ra108} (\bibinfo {year} {2016})}\BibitemShut {NoStop}%
	\bibitem [{\citenamefont {Marchant}\ \emph {et~al.}(1999)\citenamefont
		{Marchant}, \citenamefont {Callamaras},\ and\ \citenamefont
		{Parker}}]{Marchant1999}%
	\BibitemOpen
	\bibfield  {author} {\bibinfo {author} {\bibfnamefont {J.}~\bibnamefont
			{Marchant}}, \bibinfo {author} {\bibfnamefont {N.}~\bibnamefont
			{Callamaras}}, \ and\ \bibinfo {author} {\bibfnamefont {I.}~\bibnamefont
			{Parker}},\ }\href {\doibase 10.1093/emboj/18.19.5285} {\bibfield  {journal}
		{\bibinfo  {journal} {EMBO J}\ }\textbf {\bibinfo {volume} {18}},\ \bibinfo
		{pages} {5285} (\bibinfo {year} {1999})}\BibitemShut {NoStop}%
	\bibitem [{\citenamefont {FitzHugh}(1961)}]{FitzHugh1961}%
	\BibitemOpen
	\bibfield  {author} {\bibinfo {author} {\bibfnamefont {R.}~\bibnamefont
			{FitzHugh}},\ }\href {\doibase 10.1016/S0006-3495(61)86902-6} {\bibfield
		{journal} {\bibinfo  {journal} {Biophys J}\ }\textbf {\bibinfo {volume}
			{1}},\ \bibinfo {pages} {445} (\bibinfo {year} {1961})}\BibitemShut {NoStop}%
	\bibitem [{\citenamefont {Jones}(1984)}]{Jones1984}%
	\BibitemOpen
	\bibfield  {author} {\bibinfo {author} {\bibfnamefont {C.~K. R.~T.}\
			\bibnamefont {Jones}},\ }\href {\doibase 10.1090/S0002-9947-1984-0760971-6}
	{\bibfield  {journal} {\bibinfo  {journal} {T Am Math Soc}\ }\textbf
		{\bibinfo {volume} {286}},\ \bibinfo {pages} {431} (\bibinfo {year}
		{1984})}\BibitemShut {NoStop}%
	\bibitem [{\citenamefont {Lopez}\ and\ \citenamefont
		{Dawson}(2016)}]{Lopez2016}%
	\BibitemOpen
	\bibfield  {author} {\bibinfo {author} {\bibfnamefont {L.~F.}\ \bibnamefont
			{Lopez}}\ and\ \bibinfo {author} {\bibfnamefont {S.~P.}\ \bibnamefont
			{Dawson}},\ }\href {\doibase 10.1088/1478-3975/13/3/036006} {\bibfield
		{journal} {\bibinfo  {journal} {Phys Biol}\ }\textbf {\bibinfo {volume}
			{13}},\ \bibinfo {pages} {036006} (\bibinfo {year} {2016})}\BibitemShut
	{NoStop}%
	\bibitem [{\citenamefont {Tojyo}\ \emph {et~al.}(2008)\citenamefont {Tojyo},
		\citenamefont {Morita}, \citenamefont {Nezu},\ and\ \citenamefont
		{Tanimura}}]{Tojyo2008}%
	\BibitemOpen
	\bibfield  {author} {\bibinfo {author} {\bibfnamefont {Y.}~\bibnamefont
			{Tojyo}}, \bibinfo {author} {\bibfnamefont {T.}~\bibnamefont {Morita}},
		\bibinfo {author} {\bibfnamefont {A.}~\bibnamefont {Nezu}}, \ and\ \bibinfo
		{author} {\bibfnamefont {A.}~\bibnamefont {Tanimura}},\ }\href {\doibase
		10.1254/jphs.08021FP} {\bibfield  {journal} {\bibinfo  {journal} {J Pharmacol
				Sci}\ }\textbf {\bibinfo {volume} {107}},\ \bibinfo {pages} {138} (\bibinfo
		{year} {2008})}\BibitemShut {NoStop}%
	\bibitem [{\citenamefont {Smith}\ \emph {et~al.}(2009)\citenamefont {Smith},
		\citenamefont {Wiltgen}, \citenamefont {Shuai},\ and\ \citenamefont
		{Parker}}]{Smith2009b}%
	\BibitemOpen
	\bibfield  {author} {\bibinfo {author} {\bibfnamefont {I.~F.}\ \bibnamefont
			{Smith}}, \bibinfo {author} {\bibfnamefont {S.~M.}\ \bibnamefont {Wiltgen}},
		\bibinfo {author} {\bibfnamefont {J.}~\bibnamefont {Shuai}}, \ and\ \bibinfo
		{author} {\bibfnamefont {I.}~\bibnamefont {Parker}},\ }\href {\doibase
		10.1126/scisignal.2000466} {\bibfield  {journal} {\bibinfo  {journal} {Sci
				Signal}\ }\textbf {\bibinfo {volume} {2}},\ \bibinfo {pages} {ra77} (\bibinfo
		{year} {2009})}\BibitemShut {NoStop}%
	\bibitem [{\citenamefont {Marx}\ \emph {et~al.}(2001)\citenamefont {Marx},
		\citenamefont {Gaburjakova}, \citenamefont {Gaburjakova}, \citenamefont
		{Henrikson}, \citenamefont {Ondrias},\ and\ \citenamefont
		{Marks}}]{Marx2001}%
	\BibitemOpen
	\bibfield  {author} {\bibinfo {author} {\bibfnamefont {S.~O.}\ \bibnamefont
			{Marx}}, \bibinfo {author} {\bibfnamefont {J.}~\bibnamefont {Gaburjakova}},
		\bibinfo {author} {\bibfnamefont {M.}~\bibnamefont {Gaburjakova}}, \bibinfo
		{author} {\bibfnamefont {C.}~\bibnamefont {Henrikson}}, \bibinfo {author}
		{\bibfnamefont {K.}~\bibnamefont {Ondrias}}, \ and\ \bibinfo {author}
		{\bibfnamefont {A.~R.}\ \bibnamefont {Marks}},\ }\href {\doibase
		10.1161/hh1101.091268} {\bibfield  {journal} {\bibinfo  {journal} {Circu
				Res}\ }\textbf {\bibinfo {volume} {88}},\ \bibinfo {pages} {1151} (\bibinfo
		{year} {2001})}\BibitemShut {NoStop}%
	\bibitem [{\citenamefont {Ullah}\ \emph {et~al.}(2012)\citenamefont {Ullah},
		\citenamefont {Parker}, \citenamefont {Mak},\ and\ \citenamefont
		{Pearson}}]{Ullah2012}%
	\BibitemOpen
	\bibfield  {author} {\bibinfo {author} {\bibfnamefont {G.}~\bibnamefont
			{Ullah}}, \bibinfo {author} {\bibfnamefont {I.}~\bibnamefont {Parker}},
		\bibinfo {author} {\bibfnamefont {D.-O.~D.}\ \bibnamefont {Mak}}, \ and\
		\bibinfo {author} {\bibfnamefont {J.~E.}\ \bibnamefont {Pearson}},\ }\href
	{\doibase 10.1016/j.ceca.2012.04.018} {\bibfield  {journal} {\bibinfo
			{journal} {Cell Calcium}\ }\textbf {\bibinfo {volume} {52}},\ \bibinfo
		{pages} {152} (\bibinfo {year} {2012})}\BibitemShut {NoStop}%
	\bibitem [{\citenamefont {Bootman}\ \emph {et~al.}(1997)\citenamefont
		{Bootman}, \citenamefont {Berridge},\ and\ \citenamefont
		{Lipp}}]{Bootman1997a}%
	\BibitemOpen
	\bibfield  {author} {\bibinfo {author} {\bibfnamefont {M.~D.}\ \bibnamefont
			{Bootman}}, \bibinfo {author} {\bibfnamefont {M.~J.}\ \bibnamefont
			{Berridge}}, \ and\ \bibinfo {author} {\bibfnamefont {P.}~\bibnamefont
			{Lipp}},\ }\href {\doibase 10.1016/S0092-8674(00)80420-1} {\bibfield
		{journal} {\bibinfo  {journal} {Cell}\ }\textbf {\bibinfo {volume} {91}},\
		\bibinfo {pages} {367} (\bibinfo {year} {1997})}\BibitemShut {NoStop}%
	\bibitem [{\citenamefont {Marchant}\ and\ \citenamefont
		{Parker}(2001)}]{Marchant2001}%
	\BibitemOpen
	\bibfield  {author} {\bibinfo {author} {\bibfnamefont {J.~S.}\ \bibnamefont
			{Marchant}}\ and\ \bibinfo {author} {\bibfnamefont {I.}~\bibnamefont
			{Parker}},\ }\href {https://www.ncbi.nlm.nih.gov/pubmed/11226156} {\bibfield
		{journal} {\bibinfo  {journal} {EMBO J}\ }\textbf {\bibinfo {volume} {20}},\
		\bibinfo {pages} {65} (\bibinfo {year} {2001})}\BibitemShut {NoStop}%
	\bibitem [{\citenamefont {Parker}\ and\ \citenamefont {Yao}(1996)}]{Par1996}%
	\BibitemOpen
	\bibfield  {author} {\bibinfo {author} {\bibfnamefont {I.}~\bibnamefont
			{Parker}}\ and\ \bibinfo {author} {\bibfnamefont {Y.}~\bibnamefont {Yao}},\
	}\href {http://www.ncbi.nlm.nih.gov/pubmed/8815201} {\bibfield  {journal}
		{\bibinfo  {journal} {J Physiol}\ }\textbf {\bibinfo {volume} {491}},\
		\bibinfo {pages} {663} (\bibinfo {year} {1996})}\BibitemShut {NoStop}%
	\bibitem [{\citenamefont {{\O}ksendal}(2003)}]{Oksendal2003}%
	\BibitemOpen
	\bibfield  {author} {\bibinfo {author} {\bibfnamefont {B.}~\bibnamefont
			{{\O}ksendal}},\ }\href {\doibase 10.1007/978-3-642-14394-6} {\emph {\bibinfo
			{title} {{Stochastic Differential Equations}}}},\ \bibinfo {edition} {6th}\
	ed.,\ Universitext\ (\bibinfo  {publisher} {Springer Berlin Heidelberg},\
	\bibinfo {address} {Berlin, Heidelberg},\ \bibinfo {year} {2003})\ pp.\
	\bibinfo {pages} {109--112}\BibitemShut {NoStop}%
	\bibitem [{\citenamefont {Horchler}(2013)}]{Horchler2013a}%
	\BibitemOpen
	\bibfield  {author} {\bibinfo {author} {\bibfnamefont {A.~D.}\ \bibnamefont
			{Horchler}},\ }\href {https://github.com/horchler/SDETools} {\enquote
		{\bibinfo {title} {{Matlab Toolbox for the Numerical Solution of Stochastic
					Differential Equations (Version 1.2) [Source code].}}}\ } (\bibinfo {year}
	{2013})\BibitemShut {NoStop}%
\end{thebibliography}
\end{document}



\title{Supporting Material for ``A phenomenological cluster-based model of Ca\textupperscript{2+} waves and oscillations for Inositol 1,4,5-trisphosphate receptor (IP\textunderscript{3}R) channels"}


\author{Svitlana Braichenko}
\altaffiliation{Corresponding author}
\email{braichenko.svitlana@gmail.com}
\author{Atul Bhaskar}
\affiliation{Faculty of Engineering and the Environment, University of Southampton, SO17 1BJ, Southampton, UK}

\author{Srinandan Dasmahapatra}
\affiliation{School of Electronics and Computer Science, University of Southampton, SO17 1BJ, Southampton, UK}


\date{\today}



\maketitle


\section{Bifurcation analysis}

In the Subsec.~III A in the main text, we obtain qualitatively different behavior of a cluster of channels compared to R{\"{u}}diger~\cite{Rudiger2014} except for transition from the excitable to the bistable state of a cluster with an increase in $N$. Here we propose the more detailed explanation of such a transition by performing a bifurcation analysis of the two-variable system of Eqs.~(1)--(2) assuming that $c=c_d$ according to Eq.~(4). Thus, the system reads as

\begin{eqnarray}
\frac{\dif a}{\dif t} & = & f(a, h, c_d),\label{eq:a_short}\\
\frac{\dif h}{\dif t} & = & g(a, h, c_d),\label{eq:h_short}\\
c_d & = & c_{0}+\frac{1}{2}\alpha Na\Bigg\{1+\tanh\bigg[\frac{Na-1}{\epsilon}\bigg]\Bigg\},\label{eq:c}
\end{eqnarray}

\begin{figure}[h!]
	\centering{
		\includegraphics{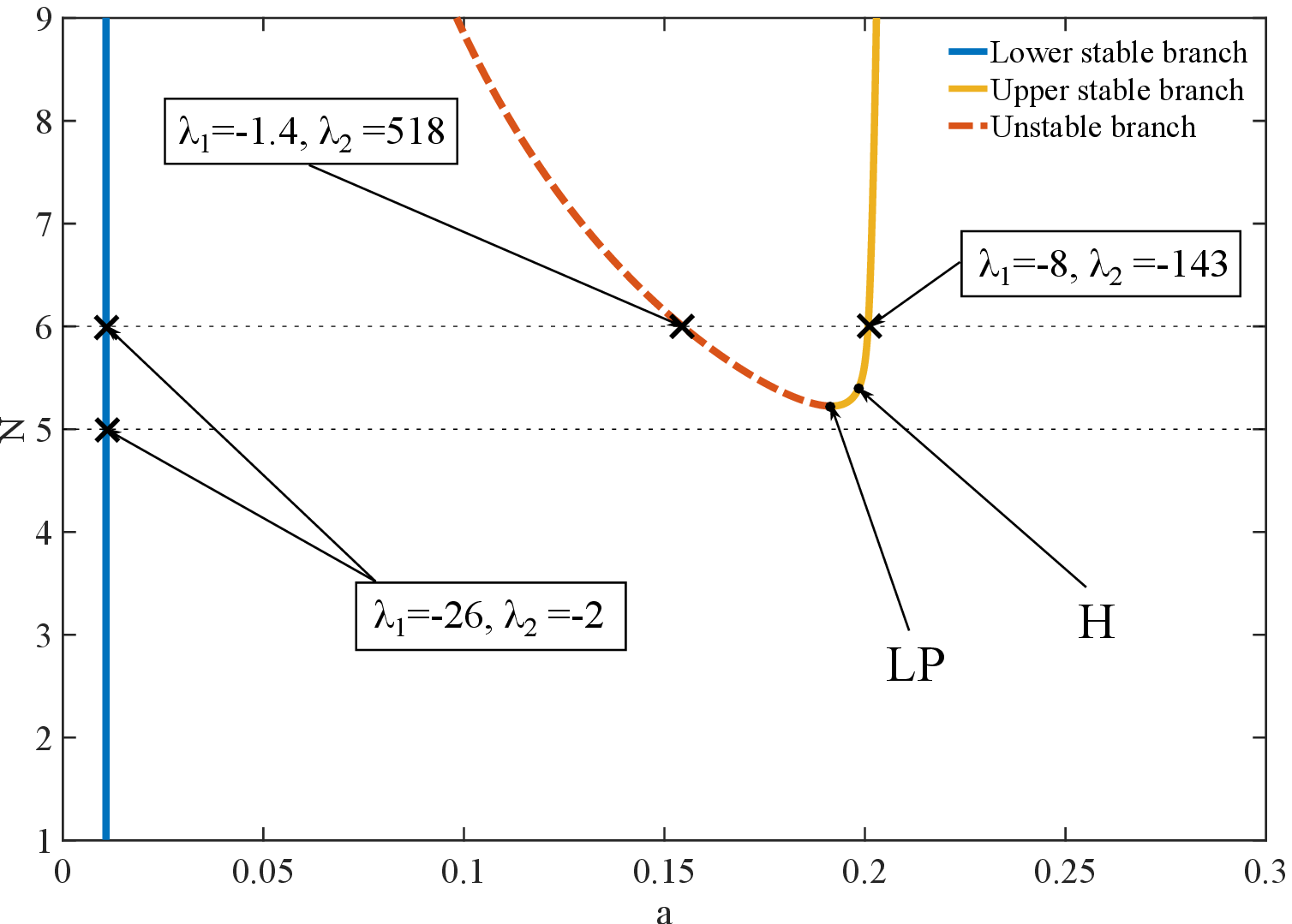}
	}
	
	\caption{ (Color online) The bifurcation diagram of the three-state system depending on the number of channels in the cluster $N$. The lower stable branch (\textit{blue solid} line) corresponds to the lower stable fixed point in Fig.~3. The eigenvalues of the Jacobian on this branch are always real and negative ($\lambda_1=-26$, $\lambda_2=-2$). The upper stable (\textit{yellow solid} line) and unstable (\textit{red dashed} line) branches occur as the consequence of the saddle-node (limit point) bifurcation marked by LP. Also, subcritical Hopf bifurcation occurs on a stable branch (marked by H). The positions of fixed points for $N=6$ correspond to the fixed points in Fig.~3(b). The stability of the branches are defined by the eigenvalues (for $N=6$ eigenvalues $\lambda_1=-8$, $\lambda_2=-143$ correspond to stable fixed point and $\lambda_1=-1.4$, $\lambda_2=518$ correspond to unstable saddle point).}	
	\label{fig:BifDiagN}	
\end{figure}

The continuation analysis of the system has been performed in order to study the stability of a cluster depending on the number of channels it comprises. We plot the bifurcation diagram of the system in Fig.~\ref{fig:BifDiagN}. Three distinctive stability branches are shown in the figure. For the clusters with $N=5$ and fewer channels only one stable branch exists. The point marked by the cross for $N=5$ channels corresponds to the stable fixed point in Fig.~3(a) in the main text. According to the lower stable (\textit{blue solid}) branch, the position of the lower fixed point does not depend on $N$ and the eigenvalues of Jacobian are always negative. With an increase in $N$ saddle-node bifurcation (or limit point marked by LP) occurs. Thus, for $N=6$ we observe three fixed points as shown in Fig.~3~(b). Notably, the eigenvalues are negative on stable (\textit{yellow solid}) and have opposite signs on the unstable saddle (\textit{red dashed}) branch. Also, subcritical Hopf bifurcation in unstable limit cycle occurs. 

Therefore, we conclude that the transition from the excitable to the bistable behavior in a cluster occurs as the result of saddle-node bifurcation (limit point LP) between $N=5$ and $N=6$ channels.

\section{Alternate oscillations} 

Here we present the extension of the results in Subsec.~III C. The light spots in the map shown in the Fig.~7~(a) and higher values of the periods in the Fig.~7~(c) correspond to the irregular behavior in Ca\textupperscript{2+} oscillations. After initial excitation passes the membrane, the oscillations emerge alternatively from the raised concentration at the bistable cluster firstly towards $x=0\,\mu$m, then to the other side towards $x=70\,\mu$m. Such behavior causes the periods of oscillations almost twice higher than in regular oscillations.     

\begin{figure}[h]
	\centering{
		\includegraphics{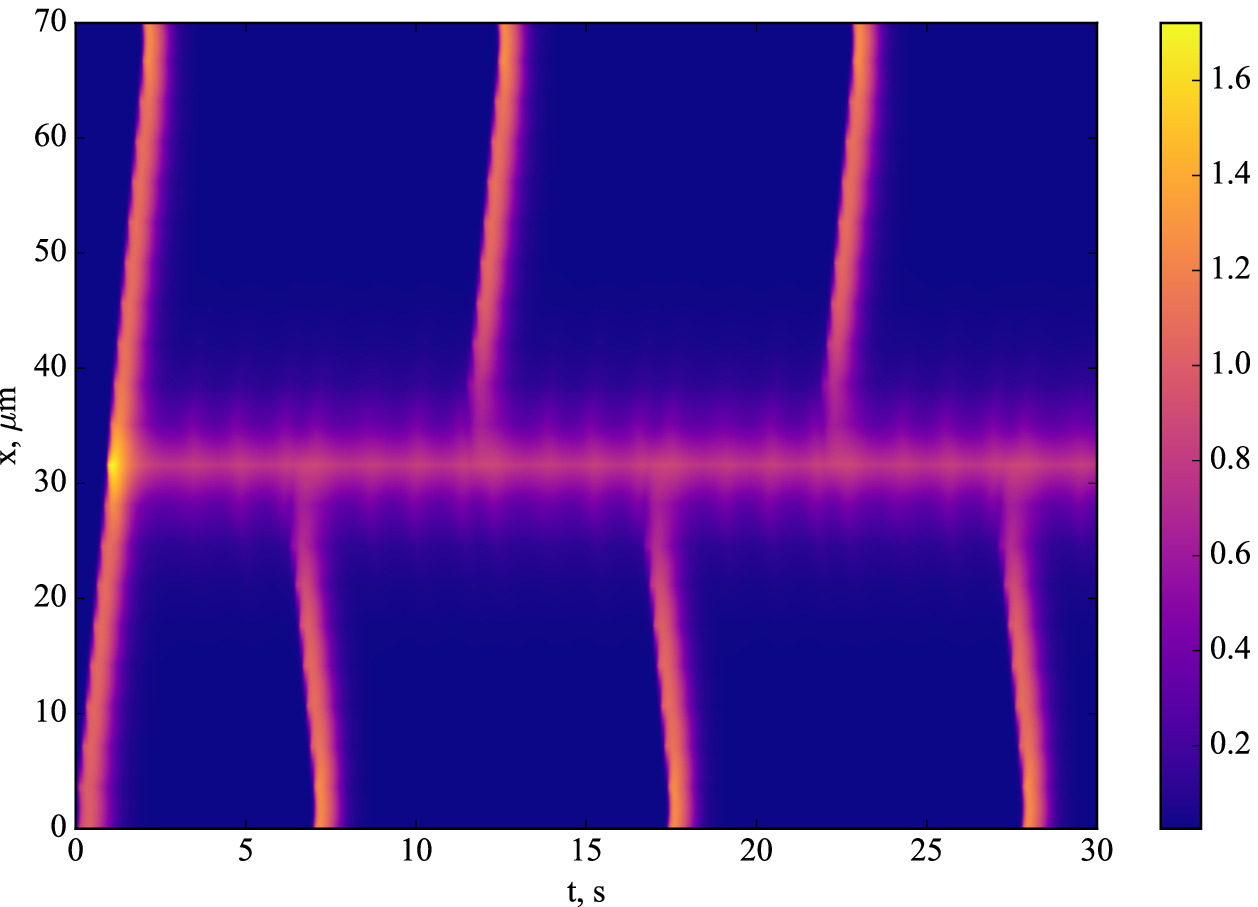}
	}
	
	\caption{ (Color online) The alternate oscillations emerging from the bistable cluster at $D=38.5\,\mu \text{m}^2/$s, $\lambda=170\,\text{s}^{-1}$.}	
	\label{fig:conc_alternate}	
\end{figure}

\section{IP3 change}

\begin{figure}[h]
	\centering{
		\includegraphics{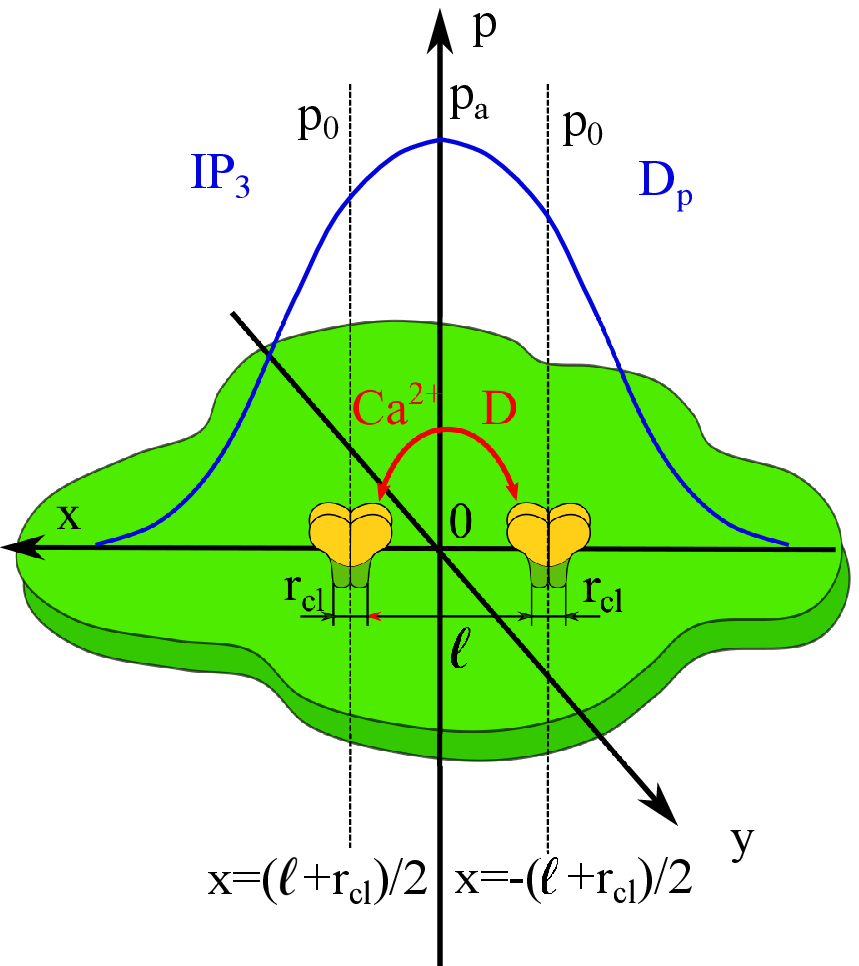}
	}
	
	\caption{ (Color online) The schematic representation of the \ip diffusion in the infinite 2D domain containing two channels.}	
	\label{fig:IP3}	
\end{figure}

We explain the termination of Ca\textupperscript{2+} release by \ip unbinding in Subsec.~III D. In this case two clusters are linked by the diffusion-like term as shown in eqs.~(10), (11). The change of the [IP\textunderscript{3}] is also governed by the diffusion. To estimate the timescales of the \ip change we solve the diffusion equation in the infinite 2D domain (Fig.~\ref{fig:IP3}). The initial shape of the \ip excitation (\textit{blue line} in Fig.~\ref{fig:IP3}) is assumed as Gaussian: 

\begin{equation}
p(x, y, 0) = p_a \operatorname{e}^{\frac{-x^2-y^2}{2 \sigma_0^2}},
\end{equation}      
the amplitude is $p_a=4.6\,\mu$M and variance $\sigma_0^2 = 1\,\mu \text{m}^2$.

The solution of the diffusion equation in infinite 2D domain is  

\begin{equation}
p(x,y,t) = \int_{-\infty}^{+\infty}{\int_{-\infty}^{+\infty}{p(\xi,\zeta,0)G(x-\xi,y-\zeta,t)\dif \xi \dif \zeta}}, \label{eq:pdiff}
\end{equation}
where the Green's function is
\begin{equation}
G(x-\xi,y-\zeta,t) = \frac{1}{4 \pi D_p t} \operatorname{e}^{-\frac{(x - \xi)^2 + (y - \zeta)^2}{4 D_p t}}.
\end{equation}

By integrating Eq.~(\ref{eq:pdiff}) we obtain the bottom case of the Eq.~(15) in the main text

\begin{equation}
p(x, y, t) = \frac{p_a \sigma_0^2}{\sigma_0^2 + 2 D_p t} \operatorname{e}^\frac{-x^2-y^2}{4 D_p t + 2 \sigma_0^2},
\end{equation}
where $D_p=10\,\mu\text{m}^2/$s. The \ip traces at each cluster (Fig.~\ref{fig:IP3} at $x=\pm(\ell + r_{cl})/2$, $y=0$) are shown as \textit{red dashed lines} in Figs.~8~(d) and (e). 

%